\documentclass[10pt,a4paper]{article}

\usepackage{amstext,amsgen,latexsym}
\usepackage{amstext,amssymb,amsfonts,latexsym}
\usepackage{theorem}
\usepackage{pifont}
\usepackage[dvips]{graphics,epsfig}

 \newcommand{\bs}{\bigskip}
 \newcommand{\ms}{\medskip}
 \newcommand{\n}{\noindent}
 \newcommand{\s}{\smallskip}
 \newcommand{\hs}[1]{\hspace*{ #1 mm}}
 \newcommand{\vs}[1]{\vspace*{ #1 mm}}

 \newcommand{\setempty}{\mathrm{\O}}
 \newcommand{\real}{\mathbb{R}}

 \newcommand{\nat}{\mathbb{N}}

 \newcommand{\complex}{\mathbb{C}}

 \newcommand{\ie}{\textrm{i.e.},\hspace*{2mm}}
 \newcommand{\eg}{\textrm{e.g.},\hspace*{2mm}}
 \newcommand{\etal}{\textrm{et al.}\hspace*{2mm}}
 \newcommand{\etalc}{\textrm{et al.}}

 \newcommand{\FF}{{\cal F}}

 \newcommand{\HH}{{\cal H}}

 \newcommand{\GG}{{\cal G}}

 \newcommand{\UU}{{\cal U}}

 \newcommand{\p}{\mathrm{P}}
 \newcommand{\np}{\mathrm{NP}}
 
 \newcommand{\rp}{\mathrm{RP}}

 \newcommand{\fp}{\mathrm{FP}}
 \newcommand{\sharpp}{\#\mathrm{P}}

 \def\bbox{\vrule height6pt width6pt depth1pt}

\theoremstyle{plain}
\theoremheaderfont{\bfseries}
 \newtheorem{theorem}{Theorem}[section]
 \newtheorem{lemma}[theorem]{Lemma}
 \newtheorem{proposition}[theorem]{Proposition}
 \newtheorem{corollary}[theorem]{Corollary}

 {\theorembodyfont{\rmfamily}
  }
 {\theorembodyfont{\rmfamily} \newtheorem{example}[theorem]{Example}}
 {\theorembodyfont{\rmfamily} }

 \newenvironment{proof}{\par \noindent
            {\bf Proof. \hs{2}}}{\hfill$\Box$ \vspace*{3mm}}

 \newenvironment{proofof}[1]{\vspace*{5mm} \par \noindent
         {\bf Proof of #1.\hs{2}}}{\hfill$\Box$ \vspace*{3mm}}

 \newcommand{\pair}[1]{\langle #1 \rangle}

\newif\ifnotesw\noteswtrue
   {\ifnotesw\marginpar[\hfill\(\top\)]{\(\top\)}\fi}%
      {\ifnotesw\marginpar[\hfill\(\bot\)]{\(\bot\)}\fi}
      
\newcommand{\mnote}[1]%
   {\ifnotesw\marginpar%
	  [{\scriptsize\begin{minipage}[t]{\marginparwidth}
	  \raggedleft#1%
		  \end{minipage}}]%
	  {\scriptsize\begin{minipage}[t]{\marginparwidth}
	  \raggedright#1%
		  \end{minipage}}%
    \fi}

\newcommand{\ignore}[1]{}

\newcommand{\holant}{\mathrm{Holant}}
\newcommand{\holantstar}{\mathrm{Holant}^{*}}

\newcommand{\sharpcsp}{\#\mathrm{CSP}}
\newcommand{\sharpcspstar}{\#\mathrm{CSP}^{*}}
\newcommand{\APreduces}{\leq_{\mathrm{AP}}}
\newcommand{\APequiv}{\equiv_{\mathrm{AP}}}

\newcommand{\DG}{{\cal DG}}
\newcommand{\NZ}{{\cal NZ}}
\newcommand{\ED}{{\cal ED}}

\newcommand{\DISJ}{{\cal DISJ}}
\newcommand{\NAND}{{\cal NAND}}
\newcommand{\EQ}{{\cal EQ}}
\newcommand{\csp}{\mathrm{csp}}

\begin{document}
\pagestyle{plain}
\setcounter{page}{1}

\begin{center}
{\Large {\bf A Dichotomy Theorem for the Approximate Counting of \s\\ 
Complex-Weighted Bounded-Degree Boolean CSPs}}\footnote{A preliminary version appeared under a slightly different title in the Proceedings of the 4th International Conference 
on Combinatorial Optimization and Applications (COCOA 2010), Lecture Notes in Computer Science, Springer, Vol.6508 (Part I), pp.285--299, Kailua-Kona, Hawaii, USA, December 18--20, 2010.} \bs\\
{\sc Tomoyuki Yamakami}\footnote{Current Affiliation: Department of Information Science, University of Fukui,   
3-9-1 Bunkyo, Fukui 910-8507, Japan} \bs\\
\end{center}


\begin{quote}
\n{\bf Abstract:} 
We determine the computational complexity of approximately counting 
the total weight 
of variable assignments for every complex-weighted Boolean constraint satisfaction problem (or CSP) with 
any number of additional unary (\ie arity $1$) constraints, 
particularly, when degrees of input 
instances are bounded from above by a fixed constant. 
All degree-$1$ counting CSPs are obviously solvable in polynomial time. 
When the instance's degree is more than two, we present a 
dichotomy theorem that classifies all counting CSPs admitting  
free unary constraints into exactly two categories.   
This classification theorem extends, to complex-weighted problems, an earlier result on the approximation 
complexity of unweighted counting  
Boolean CSPs of bounded degree. 
The framework of the proof of our theorem is based on 
a theory of signature developed from    
Valiant's holographic algorithms that can efficiently solve seemingly intractable counting CSPs. 
Despite the use of arbitrary complex weight, our proof of the classification theorem is rather 
elementary and intuitive due to an extensive use of a novel notion of limited T-constructibility.  
For the remaining degree-$2$ problems, in contrast,  
they are as hard to approximate as Holant problems, which are a generalization of counting CSPs. 

\s

\end{quote}

\section{Bounded-Degree Boolean \#CSPs}\label{sec:introduction}

Our general objective is to determine the approximation complexity\footnote{We use this term to mean the computational complexity of approximately solving a given problem.} of  constraint satisfaction problems (or CSPs) 
whose instances consist of {\em variables} (on certain domains) and 
{\em constraints}, which 
describe ``relationships'' among the variables. 
Such CSPs have found numerous applications in graph theory, database theory, and artificial intelligence as well as statistical physics.
A {\em decision CSP}, for instance, asks whether or not, 
for two given sets of variables and of constraints,  
any assignment that assigns actual values in the domain 
to the variables satisfies all the constraints simultaneously.  
The {\em satisfiability problem} (SAT) of deciding the existence of a satisfying truth assignment for a given Boolean formula is a 
typical example of the decision CSPs. 
Since input instances are often restricted to 
particular types of constraints (where a set of these 
constraints is known as a {\em 
constraint language}), it seems 
natural to parameterize CSPs  
in terms of a given set $\FF$ of allowable constraints.  
Conventionally, such a parameterized CSP is expressed 
as $\mathrm{CSP}(\FF)$.
Schaefer's \cite{Sch78} dichotomy theorem classifies all such parametrized  
$\mathrm{CSP}(\FF)$'s into exactly two categories: polynomial-time solvable problems (\ie in $\p$) and NP-complete problems, provided that $\np$ is different from $\p$. This situation highlights structural simplicity of the  $\mathrm{CS}(\FF)$'s, because all $\np$ problems, by contrast,  
fill up infinitely many categories \cite{Lad75}.  

In the course of a study of CSPs, various restrictions have been 
imposed on constraints as well as variables. 
Of all such restrictions, recently there has been a great interest in a 
particular type of restriction, of which each individual variable should not
appear more than $d$ times in the scope of all given constraints. 
The maximal number of such $d$  on any instance is called the {\em degree} of the instance. 
This degree has played a key role in a discussion of the complexity of 
CSPs; 
for instance, the {\em planar read-trice satisfiability problem}, which 
comprised of logical formulas of degree at most three,  is known to be 
$\np$-complete, while the {\em planar read-twice satisfiability problem}, whose degree is two, falls into $\p$. Those CSPs whose instances have their degrees upper-bounded are referred to as {\em bounded-degree} CSPs. 
Under the assumption that unary constraints are freely available as part of input instances, Dalmau and Ford \cite{DF03}, for example, showed that, for certain cases of $\FF$, the complexity of solving $\mathrm{CSP}(\FF)$ remains unchanged 
even if all instances are restricted to degree at most three. Notice that  
such a free use of unary constraints were frequently made in the past literature (see, \eg \cite{DF03,DGJR09,DGJ10,Fed01}) to draw stronger and more concise results.  

Apart from those decision CSPs, a {\em counting CSP} (or \#CSP, in short) asks how many variable assignments satisfy the set of given constraints. 
In parallel to Schaefer's theorem, 
Creignou and Herman \cite{CH96} gave their dichotomy theorem 
on the computational complexity of Boolean \#CSPs. Their result was 
later extended by Dyer, Goldberg, and Jerrum \cite{DGJ09} to non-negative weighted Boolean \#CSPs and then further extended  to complex-weighted Boolean \#CSPs by Cai, Lu, and Xia \cite{CLX09@}. 
Cai \etal also studied the complexity of complex-weighted Boolean \#CSPs 
whose maximal degree does not exceed three. 
Those remarkable results are meant for the computational complexity of ``exact counting.'' 
{}From a perspective of ``approximate counting,''  on the contrary,  
Dyer, Goldberg, and Jerrum \cite{DGJ10} showed a classification 
theorem on the approximate counting of the number of variable assignments for 
unweighted Boolean CSPs, parametrized by 
the choice of constraint set $\FF$, under a notion of {\em approximation-preserving (or AP) reducibility}. 
This theorem, however, is quite different from the earlier   
dichotomy theorems  for the exact-counting of Boolean CSPs; in fact, 
the theorem classifies all Boolean \#CSPs into three categories, including an intermediate level located between a class of $\p$-computable problems and a class of $\sharpp$-complete problems.  

The degree bound of input instances to \#CSPs has drawn an unmistakable picture 
in a discussion on the approximation complexity of the \#CSPs by  
Dyer, Goldberg, Jalsenius, and Richerby \cite{DGJR09}. They  
discovered the following approximation-complexity 
classification of unweighted 
Boolean \#CSPs when their degrees are further bounded. The succinct 
notation $\sharpcsp_{d}^{c}(\FF)$ used below specifies 
a problem of counting the 
number of Boolean assignments satisfying a given CSP, provided that 
(i) any unary unweighted Boolean constraint is allowed to use for 
free of charge and (ii) 
each variable appears at most $d$ times among all given constraints, including free unary constraints.  
\begin{quote}
Let $d\geq3$ and let $\FF$ be any set of unweighted Boolean constraints.  If every constraint in $\FF$ is affine, then $\sharpcsp_{d}^{c}(\FF)$ is in $\fp$.  Otherwise, if $\FF$ is included in $IM\mbox{-}conj$, then $\sharpcsp_{d}^{c}(\FF) \APequiv \#\mathrm{BIS}$. Otherwise, if $\FF\subseteq OR\mbox{-}conj$ or $\FF\subseteq NAND\mbox{-}conj$, then $\#w\mbox{-}\mathrm{HIS}_{d}\APreduces \sharpcsp_{d}^{c}(\FF) \APreduces \#w\mbox{-}\mathrm{HIS}_{kd}$. Otherwise, $\sharpcsp_{d}^{c}(\FF)\APequiv \#\mathrm{SAT}$, 
where $w$ is the width of $\FF$ and 
$k$ is a certain constant depending only on $\FF$.  
\end{quote}
Here, $IM\mbox{-}conj$, $OR\mbox{-}conj$, $NAND\mbox{-}conj$ 
are three well-defined sets of unweighted Boolean constraints, 
$\#\mathrm{SAT}$ is the counting satisfiability problem, $\#\mathrm{BIS}$ is the bipartite independent set problem, and $\#w\mbox{-}\mathrm{HIS}_{d}$ denotes the hypergraph independent set problem with hyperedge degree at 
most $d$ and width at most $w$. The notations $\APreduces$ and $\APequiv$ respectively refer to the AP-reducibility and AP-equivalence between two counting problems. 
As a special case, when $d\geq25$ and $w\geq2$, as shown in \cite{DFJ02}, 
there is no {\em fully polynomial-time randomized approximation scheme} (or FPRAS) for $\#w\mbox{-}\mathrm{HIS}_{d}$ unless $\np=\rp$.  
This  classification theorem heavily relies on the aforementioned work of Dyer \etalc~\cite{DGJ10}. 

Toward our main theorem, we first introduce a set $\ED$ of complex-weighted  constraints constructed from  unary constraints, the equality constraint, and the disequality constraint. 
Similar to the above case of Dyer \etalc, we also allow a free use of arbitrary (complex-weighted) unary constraints. 
For notational convenience, we use the notation $\sharpcspstar_{d}(\FF)$ to emphasize that all complex-weighted unary constraints are freely given.
The main purpose of this paper is to prove the following dichotomy 
theorem that classifies all $\sharpcspstar_{d}(\FF)$'s into 
exactly two categories.

\begin{theorem}\label{main-theorem}
Let $d\geq3$ be any degree bound. If $\FF\subseteq \ED$,  then $\sharpcspstar_{d}(\FF)$ belongs to 
$\fp_{\complex}$; otherwise, $\#\mathrm{SAT}_{\complex} \APreduces \sharpcspstar_{d}(\FF)$ holds. 
\end{theorem}
Here, $\#\mathrm{SAT}_{\complex}$ is a complex-weighted version of the counting satisfiability problem and $\fp_{\complex}$ is the class of polynomial-time computable complex-valued functions (see Sections \ref{sec:randomized-scheme}--\ref{sec:counting-problem} and \ref{sec:constraint-set} 
for their precise definitions).  
In contrast to the result of Dyer \etalc~\cite{DGJ10}, Theorem \ref{main-theorem} exhibits a stark difference between unweighted Boolean constraints and complex-weighted Boolean constraints, partly because of 
strong expressiveness of complex-weighted unary constraints even 
when the maximal degree of instances is upper-bounded. 

Instead of relying on the result of Dyer \etalc, 
our proof is actually based on the following dichotomy theorem 
of Yamakami \cite{Yam10}, 
who proved the theorem using a theory of {\em signature} \cite{CL08,CL07}  developed from Valiant's {\em holographic algorithms} \cite{Val06,Val08}. 
\begin{quote}
If  $\FF\subseteq \ED$,  then $\sharpcspstar(\FF)$ is in $\fp_{\complex}$. Otherwise,  $\#\mathrm{SAT}_{\complex} \APreduces \sharpcspstar(\FF)$.  
\end{quote}
To appeal to this result, we wish to claim the following key proposition, which bridges between unbounded-degree \#CSPs and bounded-degree \#CSPs.   

\begin{proposition}\label{equivalence-EQ}
For any degree bound $d$ at least $3$, $\sharpcspstar_{d}(\FF)\APequiv 
\sharpcspstar(\FF)$ holds for any set $\FF$ of complex-weighted constraints. 
\end{proposition}
{}From this proposition, Theorem \ref{main-theorem} immediately follows. 
The most part of this paper will be therefore devoted to 
proving this key proposition.  
When the degree bound $d$ equals two, on the contrary, 
$\sharpcspstar_{2}(\FF)$ is equivalent 
in approximation complexity to 
Holant problems restricted to the set $\FF$ of constraints, provided that all unary constraints are freely available, where 
{\em Holant problems} were introduced by Cai \etalc~\cite{CLX09@} 
to study a wider range of counting problems in a certain unified way.  
In the case of degree $1$, however, 
every $\sharpcspstar_{1}(\FF)$ is solvable in polynomial time. 

Our argument for complex-weighted constraints 
is obviously different from Dyer \etalc's argument for unweighted constraints and also from Cai \etalc's argument for exact counting 
using complex-valued signatures.   
While a key technique in \cite{DGJR09} is ``3-simulatability'' 
as well as ``ppp-definability,'' 
our proof argument exploits a notion of {\em limited T-constructibility}---a restricted version of 
T-constructibility developed in \cite{Yam10}. 
With its extensive use, the proof we will present in Section \ref{sec:dichotomy}  becomes quite elementary and intuitive.

\section{Preliminaries}

Let $\nat$ denote the set of all {\em natural numbers} 
(\ie non-negative integers) and $\nat^{+}$ means $\nat-\{0\}$. Similarly, $\real$ and $\complex$ denote respectively the sets of 
all {\em real numbers} and of all {\em complex numbers}. 
For succinctness, the notation $[n]$ for a number $n\in\nat^{+}$ 
expresses the integer set $\{1,2,\ldots,n\}$. The notations   
$|\alpha|$ and $\arg(\alpha)$  
for a complex number $\alpha$ denote the {\em absolute value}  and the   
{\em argument} of $\alpha$. We always assume that $\arg(\alpha)\in(-\pi,\pi]$. 
To improve readability, we often identify the ``name'' of a node 
in a given undirected graph with the ``label'' of the same node 
although there might be more than one node with the same label.  
For instance, we may call a specific node $v$ whose label is $x$ 
by ``node $x$'' as far as the node in question is clear from the context.

Hereafter, we will give brief explanations to several important concepts and notations used in the rest of the paper.

\subsection{Complex Numbers and Computability}\label{sec:complex-number}

Our core subject is the approximate computability of complex-weighted Boolean counting problems. Since such problems 
can be seen as complex-valued functions taking Boolean variables as input instances,  
we need to address a technical issue of how to handle arbitrary  
complex numbers and those complex-valued functions in an existing  framework of string-based computation. 

Our interest in this paper is not limited to so-called ``polynomial-time computable'' numbers, such as algebraic numbers, numbers expressed exactly by polynomially many bits, or numbers defined by efficiently  
generated Cauchy series \cite{Ko91}. 
Because there is no consensus of how to define ``polynomial-time computability'' of complex numbers, as done in the recent literature 
\cite{CL08,CL07,CLX09@,Yam10}, 
we wish to make our arguments in this paper independent of the definition of ``polynomial-time computable'' numbers.  
To fulfill this ambitious purpose, although slightly unconventional, 
we rather treat the complex numbers as basic ``objects'' and perform natural  ``operations'' (such as, multiplications, addition, division, subtraction, etc.)  as well as  simple  ``comparisons'' (such as, equality checking, less-than-or-equal checking, etc.) 
as basic manipulations of those numbers. 
Each of such manipulations of one or more complex numbers 
is assumed to consume only {\em constant time}.
We want to make this assumption on the constant execution time 
cause no harm in a later discussion on the computability of complex-valued functions.  
It is thus imperative to regulate all manipulations to perform only in 
a clearly described algorithmic way. 
This strict regulation guarantees that our arguments properly work in the scope of many choices of ``polynomial-time computable'' complex numbers. 

{}From a practical viewpoint, the reader may ask how we will ``describe''  arbitrary complex-valued function or, when an input instance contains 
complex numbers, how we will ``describe'' those numbers 
as a part of the input given to an algorithm in question.  
Notice that, by running a randomized algorithm within a polynomial amount of execution time, we need to distinguish only exponentially many complex numbers. Hence, those numbers may be specified by appropriately designated ``indices,'' which may be expressed in polynomially many bits. In this way, all input complex numbers, for instance, can be properly indexed when they are given as a part of each input instance, and those numbers are referred to by those indices during an execution of the algorithm. The reader is  referred to, \eg \cite[Section 4]{YY99} for a string-based treatment of arbitrary complex numbers. Indexing complex numbers also helps us view a complex-valued function as a ``map'' from Boolean variables to fixed indices of complex numbers. 

In the rest of this paper, we assume a suitable method of indexing arbitrary  complex numbers. 

\subsection{Constraints and \#CSPs}\label{sec:constraint-CSPs}

Given an undirected graph $G=(V,E)$ (where $V$ is a {\em node set} 
and $E$ is an {\em edge set})  and a node $v\in V$, an {\em incident set} 
$E(v)$ of $v$ is the set of all edges {\em incident} on $v$ (\ie $E(v) =\{w\in V|(v,w)\in E\}$), and $deg(v)$ is the {\em degree} of $v$ (\ie $deg(v) =|E(v)|$). A {\em bipartite graph} is described as a triplet of the form $(V_1|V_2,E)$, of which $V_1$ and $V_2$ respectively denote 
sets of nodes on the left-hand side and on the right-hand side of the graph and $E$ denotes a set of edges (\ie $E\subseteq V_1\times V_2$). 

Each function $f$ from $\{0,1\}^k$ to $\complex$ 
is called a {\em $k$-ary constraint} 
(or {\em signature}, in case of Holant problems), where  
$k$ is called the {\em arity} of $f$. 
Assuming the standard lexicographic order on $\{0,1\}^{k}$, we often 
express $f$ as a series of its output values, and thus it can be 
identified with an element in the space $\complex^{2^{k}}$. 
For instance, when $k=1$ and $k=2$, $f$ can be written respectively as 
$(f(0),f(1))$ and $(f(00),f(01),f(10),f(11))$. 
A constraint $f$ is {\em symmetric} if $f$'s values depend only 
on the {\em Hamming weight} of inputs.  
 When $f$ is a 
symmetric function of arity $k$, we also use a succinct notation  $f=[f_0,f_1,\ldots,f_k]$, where each $f_i$ is the value of $f$ 
on any input of Hamming weight $i$. 
As a simple example, the equality function $EQ_k$ of arity $k$ 
is expressed as $[1,0,\ldots,0,1]$ ($k-1$ zeros).  In particular, $EQ_1$ equals $[1,1]$. 
For convenience, let $\Delta_0 =[1,0]$ and $\Delta_1=[0,1]$. 
For a later use, we 
reserve the notation $\UU$ for the set of all unary (\ie arity-$1$) constraints. 

\sloppy We quickly review a set of useful notations used in \cite{Yam10}. 
Let $k\in\nat^{+}$, let $i,j\in[k]$, let $c\in\{0,1\}$, and let 
$f$ be any arity-$k$ constraint. Moreover, let $x_1,\ldots,x_k$ be $k$ Boolean variables. 
{\em Pinning} is a method of constructing a new constraint 
$f^{x_i=c}$ from $f$, where 
$f^{x_i=c}$ is the constraint defined by  $f^{x_i=c}(x_1,\ldots,x_{i-1},x_{i+1},\ldots,x_k) = f(x_1,\ldots,x_{i-1},c,x_{i+1},\ldots,x_k)$.  
In contrast, {\em projection} is a way of building a new constraint 
$f^{x_i=*}$  that is defined as  $f^{x_i=*}(x_1,\ldots,x_{i-1},x_{i+1},\ldots,x_k) = \sum_{x_i\in\{0,1\}}f(x_1,\ldots,x_{i-1},x_i,x_{i+1},\ldots,x_k)$. 
When $i\neq j$, the notation $f^{x_j=x_i}$ denotes the constraint 
defined as  $f^{x_j=x_i}(x_1,\ldots,x_{j-1},x_{j+1},\ldots,x_k) = f(x_1,\ldots,x_{j-1},x_{i},x_{j+1},\ldots,x_k)$.  
To {\em normalize} $f$ means that we choose an appropriate constant $\lambda\in\complex-\{0\}$ and then construct a new constraint $\lambda\cdot f$ from $f$, where $\lambda\cdot f$ denotes the constraint $g$ defined as $g(x_1,\ldots,x_k) = \lambda\cdot f(x_1,\ldots,x_k)$.  
When $g_1$ and $g_2$ share the same input-variable sequence,    
$g_1\cdot g_2$ denotes the constraint $h$ defined as $h(x_1,\ldots,x_k) = g_1(x_1,\ldots,x_k) g_2(x_1,\ldots,x_k)$. 
By  extending these notations naturally, we abbreviate, \eg $(f^{x_1=0})^{x_2=1}$ as $f^{x_1=0,x_2=1}$ and $(f^{x_1=*})^{x_2=*}$ as $f^{x_1=*,x_2=*}$.  

For each set $\FF$ of constraints, a {\em complex-weighted Boolean 
\#CSP}, succinctly denoted $\sharpcsp(\FF)$, is a counting problem whose input instance is a finite set $\Omega$ of ``elements'' of 
the form $\pair{h,(x_{i_1},x_{i_2},\ldots,x_{i_k})}$, where $h:\{0,1\}^k\rightarrow\complex$ is in $\FF$ and  $x_{i_1},x_{i_2},\ldots,x_{i_k}$ are some of $n$ Boolean variables $x_1,x_2,\ldots,x_n$ (\ie $i_1,\ldots,i_k\in[n]$), and $\sharpcsp(\FF)$ asks to compute the value 
$
\csp_{\Omega}= \sum_{\sigma} \prod_{\pair{h,x}\in H} h(\sigma(x_{i_1}),\sigma(x_{i_2}),\ldots,\sigma(x_{i_k})), 
$ 
where $x=(x_{i_1},x_{i_2},\ldots,x_{i_k})$ and  $\sigma:\{x_1,x_2,\ldots,x_n\}\rightarrow\{0,1\}$ ranges over the set of all {\em variable assignments}. 
To improve readability, we often omit the set notation and express, \eg $\sharpcsp(f,g,\FF,\GG)$ to mean $\sharpcsp(\{f,g\}\cup \FF\cup\GG)$.  
Since we always admit arbitrary unary constraints for free of charge, we briefly write $\sharpcspstar(\FF)$ instead of $\sharpcsp(\FF,\UU)$. 

{}From a different viewpoint, 
an input instance $\Omega$ to $\sharpcsp(\FF)$ can be stated as a triplet 
$(G,X|\FF',\pi)$, 
which consists of a finite undirected bipartite graph $G=(V_1|V_2,E)$, a variable set $X=\{x_1,x_2,\ldots,x_n\}$,   
a {\em finite} subset $\FF'$ of $\FF$, and a labeling function $\pi:V_1\cup V_2 \rightarrow X\cup\FF'$, where $\pi(V_1)=X$ and $\pi(V_2)\subseteq\FF'$. In this graph representation,  
the label of each node $v$ in $V_1$ is distinct variable $x_i$ in $X$, each node $w$ in $V_2$ has  constraint $h$ in $\FF'$ as its label, and an edge $e$ in $E$ incident on both nodes $v$ and $w$ indicates 
that the constraint $h$ takes the variable $x_i$ (as part of its input variables). Such labeling of constraints is formally given by the labeling function $\pi$. 
For simplicity, $\pi(v)$ is written as $f_{v}$. 
To emphasize this graph representation, we intend to call $\Omega=(G,X|\FF',\pi)$ a {\em constraint frame} for $\sharpcsp(\FF)$ \cite{Yam10}.  
The use of the notion of constraint frame makes it possible to discuss a counting problem $\sharpcsp(\FF)$  in a general framework of Holant problem (on a Boolean domain) \cite{CLX09@}, which will be given in the next subsection. 

For each input instance $\Omega=(G,X|\FF',\pi)$ given to $\sharpcspstar(\FF)$, the {\em degree} of the instance $\Omega$ is the greatest number of times that any variable appears among its constraints in $\FF'$; in other words, the maximum degree of any node that appears on the left-hand side of the  bipartite graph $G$.   
For any positive integer $d$, we write $\sharpcspstar_{d}(\FF)$ for the restriction of $\sharpcspstar(\FF)$ to instances of degree at most $d$. 

\subsection{Holant Problems}\label{sec:Holant-problem}

In a Holant framework, ``(complex-weighted) constraints'' are always referred to as  ``signatures.'' 
For our convenience, we often use these two words interchangeably.  
Now, we will follow the terminology developed in \cite{CL08,CL07}.
A {\em Holant problem} $\holant(\FF)$ (on a Boolean domain) takes an input instance, called 
a {\em signature grid} $\Omega =(G,\FF',\pi)$, composed of a finite undirected graph $G=(V,E)$,  
a finite set $\FF'\subseteq \FF$, and a labeling function $\pi:V \rightarrow\FF'$, where each node $v\in V$ is labeled by a signature   $\pi(v): \{0,1\}^{deg(v)}\rightarrow\complex$. 
We often use the notation $f_v$ for $\pi(v)$.  
Instead of variable assignments used for $\sharpcsp(\FF)$'s, 
here we use ``edge assignments.'' 
We denote by $Asn(E)$ the set of all edge assignments $\sigma: E\rightarrow \{0,1\}$. The Holant problem asks to compute the value 
$ 
\holant_{\Omega} = \sum_{\sigma\in Asn(E)} \prod_{v\in V}f_{v}(\sigma|E(v)), 
$     
where $\sigma|E(v)$ denotes the sequence   $(\sigma(w_1),\sigma(w_2),\cdots,\sigma(w_k))$ of bits if $E(v)=\{w_1,w_2,\ldots,w_k\}$, sorted in a certain pre-determined order (depending only on $f_v$). 
A {\em bipartite Holant problem} $\holant(\FF_1|\FF_2)$ is a variant of Holant problem, defined as follows. 
An input instance is a {\em bipartite signature grid} $\Omega =(G,\FF'_1|\FF'_2,\pi)$ consisting of a finite undirected bipartite graph $G=(V_1|V_2,E)$, 
two finite sets $\FF'_1\subseteq \FF_1$ and $\FF'_2\subseteq \FF_2$, and a labeling function $\pi:V_1\cup V_2\rightarrow\FF_1'\cup\FF'_2$ satisfying 
that  $\pi(V_1)\subseteq \FF'_1$ and $\pi(V_2)\subseteq \FF'_2$. 

Exploiting a direct connection between \#CSPs and Holant problems,  it is useful to view 
$\sharpcsp(\FF)$ as a special case of bipartite Holant problem by the following translation: 
a constraint frame $\Omega=(G,X|\FF',\pi)$ for $\sharpcsp(\FF)$ with $G=(V_1|V_2,E)$ is modified into  
a signature grid $\Omega'=(G',\{EQ_k\}_{k\geq1}|\FF',\pi')$ that is obtained as follows.  The graph $G'$ is obtained from $G$ by replacing 
 the variable label of any degree-$k$ node $v$ in $V_1$  
by the arity-$k$ equality function 
$EQ_k$.  It is not difficult to see that any edge assignment that assigns $0$ ($1$, resp.) to all edges incident on this node $v$ uniquely  
substitutes a variable assignment giving $0$ ($1$, resp.) to the node $v$ for $\sharpcsp(\FF)$.  The labeling function $\pi'$ is defined accordingly. 
In terms of Holant problems,  $\sharpcsp(\FF)$ is just another name for $\holant(\{EQ_{k}\}_{k\geq1}|\FF)$.  
Similarly, $\sharpcspstar(\FF)$ coincides with 
$\holant(\{EQ_{k}\}_{k\geq1}|\FF,\UU)$.  Moreover, for each degree bound $d\geq1$,  $\sharpcspstar_{d}(\FF)$ is identified with $\holant(\{EQ_k\}_{k\in [d]}|\FF,\UU)$. 

\subsection{{FP}$_{\complex}$ and AP-Reductions}\label{sec:randomized-scheme}

Following the way we handle complex numbers (see Section \ref{sec:complex-number}),  a complex analogue of $\fp$, denoted  
$\fp_{\complex}$, is naturally defined as the set of all 
functions, mapping strings to $\complex$, which can be computed deterministically in time polynomial in the lengths of input strings, where ``strings'' are finite sequences of symbols chosen from (nonempty finite) alphabets.  
Let $\Sigma$ be any alphabet and let $F$ be any function mapping from $\Sigma^*$ (\ie the set of all strings over $\Sigma$) to $\complex$. 
A {\em randomized approximation scheme} (or RAS) for 
$F$ is a randomized algorithm (equipped with a coin-flipping mechanism) that takes a standard input $x\in\Sigma^*$ together with an {\em error tolerance parameter} $\varepsilon\in(0,1)$ and outputs values $w$ in $\complex$ with probability at least $3/4$ for which  
$
2^{-\epsilon} \leq \left| {w}/{F(x)} \right| \leq 2^{\epsilon}$ and  
$ 
\left| \arg\left( {w}/{F(x)} \right) \right|\leq \epsilon, 
$  
provided that, whenever $F(x)=0$, we always demand  $w=0$. See \cite[Lemma 9.2]{Yam10} for usefulness of this definition. 

Given two functions $F$ and $G$, a {\em polynomial-time  approximation-preserving reduction} (or {\em AP-reduction}) from $F$ to $G$ is a randomized algorithm $M$ that takes a pair $(x,\varepsilon)\in\Sigma^*\times(0,1)$ as input instance,  
uses an arbitrary RAS $N$ for $G$ as {\em oracle}, 
and satisfies the following conditions:
(i) $M$ is still a valid RAS for $F$;  
(ii) every oracle call made by $M$ is of the form $(w,\delta)\in\Sigma^*\times(0,1)$ with $\delta^{-1}\leq poly(|x|,1/\varepsilon)$ and its answer is the outcome of $N$ on $(w,\delta)$, provided that any complex number included in this string $w$ should be completely ``specified'' (see Section \ref{sec:complex-number}) by $M$; and (iii) the running time of $M$ is bounded from above by a certain polynomial in $(|x|,1/\varepsilon)$, not depending on the choice of $N$. In this case, we write $F\APreduces G$ and we also say that $F$ is {\em AP-reducible} to $G$ (or $F$ is {\em AP-reduced} to $G$). If both $F\APreduces G$ and $G\APreduces F$ hold, then $F$ and $G$ are 
said to be {\em AP-equivalent} and we write $F\APequiv G$.  
 
The following lemma, whose proof is straightforward and left to the reader, is useful in later sections.

\begin{lemma}\label{AP-properties}
Let $\FF,\GG,\HH$ be any three constraint sets and let $e,d\in\nat^{+}$.  
\begin{enumerate}\vs{-2}
\item If $d\leq e$, then 
$\sharpcspstar_{d}(\FF)\APreduces \sharpcspstar_{e}(\FF)$.  
\vs{-2}
\item If $\FF\subseteq \GG$, then 
$\sharpcspstar_{d}(\FF)\APreduces \sharpcspstar_{d}(\GG)$. 
\vs{-2}
\item If $\sharpcspstar_{d}(\FF)\APreduces \sharpcspstar_{d}(\GG)$ and $\sharpcspstar_{d}(\GG)\APreduces \sharpcspstar_{d}(\HH)$, then $\sharpcspstar_{d}(\FF)\APreduces \sharpcspstar_{d}(\HH)$.
\end{enumerate}
\end{lemma}

\subsection{Counting Problem {\#SAT}$_{\complex}$}\label{sec:counting-problem}

We briefly describe the counting problem    
 $\#\mathrm{SAT}_{\complex}$, introduced in \cite{Yam10}, 
which has appeared in Section \ref{sec:introduction}.  
For the proof of our main theorem, since our proof heavily relies on \cite{Yam10}, there is in fact no need of knowing any structural property of this  counting problem; however, the interested reader is referred to \cite{Yam10} for its properties and connections to other counting problems. 

A complex-weighted version of the counting satisfiability problem 
($\#\mathrm{SAT}$),  denoted  
$\#\mathrm{SAT}_{\complex}$, is induced naturally from $\#\mathrm{SAT}$  
as follows. Let $\phi$ be any propositional formula and let $V(\phi)$ 
denote the set of all Boolean variables appearing in $\phi$. 
In addition, let $\{w_x\}_{x\in V(\phi)}$ be any series of 
{\em node-weight functions} 
$w_{x}:\{0,1\}\rightarrow\complex-\{0\}$ for each variable $x$ in $V(\phi)$. 
Given the input pair $(\phi,\{w_x\}_{x\in V(\phi)})$, $\#\mathrm{SAT}_{\complex}$ outputs the sum of all weights 
$w(\sigma)$ for truth assignments $\sigma$ 
satisfying $\phi$, where $w(\sigma)$ denotes the product of all  values $w_{x}(\sigma(x))$ over all variables $x\in V(\phi)$. 
Since $\#\mathrm{SAT}$ is a special case of $\#\mathrm{SAT}_{\complex}$, 
it naturally holds that 
$\#\mathrm{SAT}\APreduces \#\mathrm{SAT}_{\complex}$. 

\section{Special Constraint Sets}\label{sec:constraint-set}

We treat a {\em relation} of 
arity $k$ as both a subset of $\{0,1\}^k$ 
and a function mapping $k$ Boolean variables to $\{0,1\}$. 
{}From this duality, we often utilize the following ``functional'' notation: 
for every $x\in\{0,1\}^k$, $R(x)=1$ ($R(x)=0$, resp.) iff $x\in R$ ($x\not\in R$, resp.). 
The {\em underlying relation} $R_f$ of a constraint $f$ of arity $k$ is 
the set $\{x\in\{0,1\}^k\mid f(x)\neq0\}$. A constraint $f$ is called {\em non-zero} if $f(x)\neq0$ for all inputs $x\in\{0,1\}^k$. Note that, for any constraint $f$, there exists a {\em non-zero} constraint $g$ 
for which $f = R_f\cdot g$, where $R_f$ is viewed as a Boolean function. 
This fundamental property will be frequently 
used in the subsequent sections. 

In this paper, we use the following special relations: 
$XOR=[0,1,0]$, $Implies=(1,1,0,1)$,  
 $OR_{k}=[0,1,\ldots,1]$ ($k$ ones),  
$NAND_{k}=[1,\ldots,1,0]$ ($k$ ones), and 
$EQ_{k}=[1,0,\ldots,0,1]$ ($k-1$ zeros), where $k\in\nat^{+}$.  
Slightly abusing notations, we let the notation $EQ$ ($OR$ and $NAND$, resp.) 
refer to the equality function ($OR$-function and $NAND$-function, resp.) of {\em arbitrary arity larger than one}. This notational convention is quite useful when we do not want to specify its arity. 


Moreover, we use the following two sets of relations. 
A relation $R$ is in $DISJ$ ($NAND$, resp.)  if it equals a    
product of a {\em positive} number of relations of the forms 
$OR_k$ ($NAND_k$, resp.), $\Delta_0$, and $\Delta_1$, 
where $k\geq2$ (slightly different from $OR\mbox{-}conj$ and 
$NAND\mbox{-}conj$ in \cite{DGJR09}). Notice that the empty relation ``$\setempty$'' is  in $DISJ\cup NAND$. Next, we introduce six sets of constraints, the first four of which were defined in \cite{Yam10}. 

\begin{enumerate}\vs{-1}
\item Recall that $\UU$ denotes the set of all unary constraints. 
\vs{-2}
\item Let $\NZ$ denote the set of arbitrary non-zero constraints.
\vs{-2}
\item Let $\DG$ denote the set of all constraints $f$  
that are expressed as products of unary constraints, each of 
which is applied to a different variable of $f$. 
Every constraint in $\DG$ is called {\em degenerate}. In particular, $\UU$ is included in $\DG$. The underlying relation of any degenerate constraint  is also degenerate; however, the converse is not true in general. 
\vs{-2}
\item Let $\ED$ 
denote the set of constraints expressed as products of 
unary constraints, the binary equality $EQ_2$, and 
the binary disequality $XOR$. Clearly, $\DG\subseteq \ED$ holds. The name ``$\ED$'' refers to its key components of ``equality'' and ``disequality.''  
\vs{-2}
\item Let $\DISJ$ be the set of all constraints $f$ for which $R_f$ 
is in $DISJ$.   
\vs{-2}
\item Let $\NAND$ be the set of all constraints $f$ for which 
$R_f$ belongs to $NAND$.   
\end{enumerate}\vs{-1}

For later convenience, we list a simple characterization of 
binary constraints in $\DG$.

\begin{lemma}\label{basic-DG-IMP}
Let $f$ be any binary constraint $f=(a,b,c,d)$ with 
$a,b,c,d\in\complex$. It holds that $f\not\in \DG$ iff $ad\neq bc$.  
\end{lemma}

\begin{proof}
Let $f=(a,b,c,d)$ with $a,b,c,d\in\complex$. First, assume that $f$ is degenerate. Since $f\in\DG$, there are four constants $x,y,z,w\in\complex$ such that $f(x_1,x_2) = [x,y](x_1)\cdot[z,w](x_2)$ holds for every vector $(x_1,x_2)\in\{0,1\}^2$. This implies $f=(xz,xw,yz,yw)$. Since $f$ equals $(a,b,c,d)$, we obtain $ad = xyzw =bc$, as required. Next, we assume that $ad=bc$. There are three cases to examine separately. 

(i) Consider the case where $a=0$. By our assumption, either $b=0$ or $c=0$ holds. If $b=0$, then it holds that $f(x_1,x_2) = [0,1](x_1)\cdot[c,d](x_2)$; thus, $f$ is degenerate. Similarly, when $c=0$, we obtain $f(x_1,x_2) = [b,d](x_1)\cdot[0,1](x_2)$ and thus $f$ is degenerate. 

(ii) The case where $d=0$ is similar to Case (i). 

(iii) Finally, assume that $ad\neq0$. Obviously, $bc\neq0$ holds since $ad=bc$. Let us define  $y=\frac{b}{a}=\frac{d}{c}$. Since $b=ay$ and $d=cy$, it instantly follows that $f(x_1,x_2) =[a,c](x_1)\cdot [1,y](x_2)$. {}From this equality, we conclude that $f$ is degenerate. 
\end{proof}

A key idea of \cite{Yam10} is a certain form of 
``factorization'' of a target constraint. 
For each constraint $f$ in $\ED$, for instance,  
its underlying relation $R_f$ can be expressed by a product 
$R_f = g_1\cdot g_2\cdots g_{m}$, where 
each constraint $g_i$ is one of the following forms:  
$u(x)$, $EQ_2(x,y)$, and $XOR(x,y)$ 
(where $x$ and $y$ may be the same), where $u$ is an arbitrary unary constraint. This indicates that $f$ is 
``factorized'' into {\em factors}: $g_1,\ldots,g_m$ (which always 
include the information on input variables).  
The list $L =\{g_1,g_2,\ldots,g_{m}\}$ of all such factors  
is succinctly called a {\em factor list} for $R_f$. 

In our later argument, 
factor lists will play an essential role. Let us introduce a  
notion---an or-distinctive list---for each constraint in $\DISJ$. 
Associated with a constraint $f$ in $\DISJ$, let us consider a list $L$ 
of all factors of the form $\Delta_0(x)$, $\Delta_1(x)$, and $OR_d(x_{i_1},\ldots,x_{i_k})$, that characterizes $R_f$. This factor list $L$ is 
called {\em or-distinctive} if (i) no variable appears more than once in each $OR$ in $L$, (ii) no two factors $\Delta_{c}$ ($c\in\{0,1\}$) and $OR$ in $L$ share the same variable, 
(iii) no $OR$'s variables form a subset of any other's (when ignoring the variable order), and (iv) every $OR$ in $L$ has at least two variables. 
For each constraint in $\NAND$, we obtain a similar notion of {\em nand-distinctive list} by replacing $OR$s with $NAND$s. 

The following lemma is fundamentally the same as \cite[Lemma 3.2]{DGJR09} for Boolean constraints.

\begin{lemma}
For any constraint $f$ in $\DISJ$, there exists a unique 
or-distinctive list of all factors of $R_f$. The same holds for nand-distinctive lists and $\NAND$.
\end{lemma}

\begin{proof}
Let $f$ be any $k$-ary constraint in $\DISJ$ and let $L$ be any factor list for  $R_f$ with the condition 
that each factor in $L$ has one of the following  forms: $\Delta_0(x)$, $\Delta_1(x)$, and $OR_d(x_{i_1},\ldots,x_{i_d})$, 
where $d\geq2$ and $i_1,\ldots,i_d\in[k]$. 
Now, let us consider the following procedure that transforms $L$ into another factor list, which becomes or-distinctive. For ease of the description of this procedure, we assume that, during the procedure, whenever all variables are completely deleted from an argument place of any factor $g$ in $L$, this $g$ is automatically removed from the list $L$, since $g$ is no longer a valid constraint.  
Moreover, if there are two exactly the same factors (with the same series of input variables), then exactly one of them is automatically deleted from $L$. Finally, since $OR_1$ equals $\Delta_1$, any factor $OR_1(x)$ in $L$ is automatically replaced by $\Delta_1(x)$.   

(i) For each factor $OR_d$ in $L$, 
if a variable $x$ appears more than once in its argument place, 
then we delete the second occurrence of $x$ from the argument place. This  deletion 
causes this $OR_d$ to shrink to an $OR_{d-1}$. 
Now, we assume that every factor $OR$ in $L$ has no duplicated variables. (ii) If two factors $OR_d$  
and $\Delta_1$ in $L$ share the same variable, say, $x$, 
then we remove this $OR_d$ from $L$. This removal 
is legitimate  because this $OR_d$ is clearly redundant. 
(iii) If two factors $OR_d$ and $\Delta_0$ in $L$ 
share the same variable $x$, 
then we delete $x$ from any argument places of {\em all} $OR$s in $L$. 
This process is also legitimate, because $x$ is pinned down to $0$ by $\Delta_0(x)$ and it does not contribute to the outcome of $OR$s.  
It is not difficult to show that the list obtained from $L$ by executing this procedure is indeed or-distinctive. 

To complete the proof, we will show the uniqueness of any or-distinctive  list for $R_f$. Assume that $L_1$ and $L_2$ are two distinct or-distinctive lists of all factors of $R_f$. 
Henceforth, we intend to show that $L_1\subseteq L_2$. For simplicity, let 
$X_0=\{x_1,\ldots,x_k\}$ denote the set of all variables that do not appear in any factor of the form $\Delta_c$ ($c\in\{0,1\}$) in $L_1$. 
We note that any factor $\Delta_c$ in $L_2$ takes no variable in $X_0$ because, otherwise, $L_1$ and $L_2$ must define two different relations, a contradiction against our assumption that $L_1$ and $L_2$ are factor lists for the same relation $R_f$. Toward our goal, we need to prove two claims. 

First, we claim that all factors of the form $\Delta_{c}$ ($c\in\{0,1\}$) in $L_1$ belong to $L_2$. Assume otherwise; that is, there is a factor $\Delta_{c}(x)$ that appears in $L_1$ but not in $L_2$.  Notice that $x$ should appear in a certain factor in $L_2$. 
If the factor $\Delta_{1-c}(x)$ is present in $L_2$, then $L_1$ and $L_2$ should define two different relations, a clear contradiction. Hence, $L_2$ does not contain $\Delta_{1-c}(x)$. Since $x$ cannot appear in both $\Delta_0$ and $\Delta_1$ in $L_2$,  $x$ must appear in a certain 
$OR$, say, $h$ of arity $m$ in $L_2$. Since $L_2$ is an or-distinctive list, $m\geq2$ follows. Let us choose a variable assignment $a$ to $x$ satisfying $\Delta_{c}(a)=0$. By choosing another assignment $b\in\{0,1\}^{m-1}$ appropriately, we can force $h(a,b)=1$. This is a clear contradiction.  

Next, we claim that all $OR$s in $L_1$ are also in $L_2$. Toward a contradiction, we assume that (after appropriately permuting variable indices) 
$g(x_1,\ldots,x_d)$ is an $OR_d$ in $L_1$ but not in $L_2$. 
Let $X=\{x_1,\ldots,x_d\}$. By the or-distinctiveness, any other $OR$ in $L_1$ should contain at least one variable in $X_0-X$. 
We need to examine the following two cases separately. 
(1)' Assume that there exist an index $m\in[d-1]$ and a factor $h$ of the form $OR_{m}$ (or $\Delta_1$ if $m=1$) in $L_2$ satisfying that all variables of 
$h$ are in $X$. Since $m<d$, 
we obtain both $h(0^m)=0$ and $g(0^m,1^{d-m})=1$. This is a contradiction.
(2)' Assume that every factor $h$ of the form $OR$ in $L_2$ contains at least one variable in $X_0-X$. Clearly,  it holds that 
$g(0^d)=0$ and $h(a,b)=1$, where $a$ and $b$ are respectively 
appropriate nonempty portions of $0^d$ and $1^{k-d}$. This also leads to a  contradiction. Therefore, $g$ should belong to $L_2$. 

In the end, we conclude that $L_1\subseteq L_2$. Since we can prove by symmetry that $L_2\subseteq L_1$, this yields the equality $L_1=L_2$, and thus we establish the uniqueness of an or-distinctive list for $R_f$.  
The case for $\NAND$ can be similarly treated.   
\end{proof}

\section{Limited T-Constructibility}\label{sec:T-reduction}

A technical tool used for an analysis of \#CSPs in \cite{Yam10} is the notion of {\em T-constructibility}, which asserts that a given constraint can be systematically ``constructed''  by applying certain specific operations recursively, starting from a finite set of target constraints.   
Such a construction directly corresponds to a modification 
of bipartite graphs in constraint frames.  
Since our target is bounded-degree \#CSPs, 
we rather use its weakened version.  


Now, we introduce our key notion of {\em limited T-constructibility}, 
which will play a central role in our later arguments toward 
the proof of the main theorem.  
Let $f$ be any constraint of arity $k\geq1$ and let $\GG$ be 
any {\em finite} constraint set. 
We say that an undirected bipartite graph $G=(V_1|V_2,E)$ (implicitly with a labeling function $\pi$) {\em represents} $f$ if $V_1$ consists only of $k$ nodes labeled $x_1,\ldots,x_k$, which may have a certain number of dangling\footnote{A {\em dangling edge} is obtained from an edge by deleting exactly one end of the edge. These dangling edges are treated as ``normal'' edges. Therefore, the degree of a node should count dangling 
edges as well.} edges, and $V_2$ contains only a node labeled 
$f$, to whom every node $x_i$ is adjacent. 
As noted before, we write $f_{w}$ for $\pi(w)$. We also say that 
$G$ {\em  realizes} $f$ by $\GG$ if the following four 
conditions are met: (i) $\pi(V_2)\subseteq \GG$, 
(ii) $G$ contains at least $k$ 
nodes labeled $x_1,\ldots,x_k$, possibly together with nodes 
associated with other variables, say, $y_1,\ldots,y_m$; namely, 
$V_1=\{x_1,\ldots,x_k,y_1,\ldots,y_m\}$ (by identifying a node name with its variable label),  
(iii)  only the nodes $x_1,\ldots,x_k$ may have dangling edges, and 
(iv) $f(x_1,\ldots,x_{k}) = \lambda \sum_{y_1,\ldots,y_m\in\{0,1\}} \prod_{w \in V_2} f_{w}(z_1,\ldots,z_d)$, where $\lambda\in\complex-\{0\}$ and $\{z_1,\ldots,z_d\}$ is a subset of $V_1$.  

\begin{example}\label{realize-g}
Here, we give a useful example of an undirected bipartite graph that realizes a constraint $g$ of particular form: (*)  $g(x_1,x_2) =\sum_{y\in \{0,1\}} f(x_1,y)u(y)f(y,x_2)$. Corresponding to this equation (*), 
we construct the following graph, denoted $G^{[f,u]}$. This graph is composed of three nodes labeled $x_1,x_2,y$ on its left-hand side 
and two nodes $v_1$ and $v_2$ labeled $f$ as well as a node $w$ labeled $u$ on the right-hand side. The graph has an edge set $\{(x_1,v_1), (y,v_1),(y,w),(x_2,v_2),(y,v_2)\}$. 
Since this graph $G^{[f,u]}$ faithfully  
reflects the above equation (*), it is 
not difficult to check that Condition (iv) of the definition of realizability is satisfied. Therefore, $G^{[f,u]}$ realizes $g$ by $\{f,u\}$.
\end{example}

Let $d\in\nat$ be any index. We write 
$f\leq_{con}^{+d}\GG$ if the following conditions hold: 
for any number $m\geq 2$ and for any graph $G$ representing $f$ with distinct variables $x_1,\ldots,x_k$ whose node degrees are  at most $m$, there exists another graph $G'$ such that (i) $G'$ realizes   
$f$ by $\GG$, (ii) $G'$ has the same dangling edges  as $G$ does, 
(iii) the nodes labeled $x_1,\ldots,x_k$ have degree at most $m+d$, and (iv) 
all the other nodes on the left-hand side of $G'$ have degree at most $\max\{3,d\}$.   
In this case, we loosely say that $f$ is {\em limited T-constructible} 
from $\GG$. 
The constraint $g$ in Example \ref{realize-g} is limited T-constructible from $\{f,u\}$. More precisely, since $G^{[f,u]}$ contains the node $y$ of degree $3$, $g\leq_{con}^{+0}\{f,u\}$ holds. Although the above definition is general enough, in this paper, we are interested only in the case where   $0\leq d\leq 1$. 

We will see another example. 

\begin{example}\label{pinning-T}
Let $f$ and $g$ be any two constraints. If $f$ is obtained from $g$ 
by pinning $g$, then $f\leq_{con}^{+0}\{g,\Delta_0,\Delta_1\}$ holds. 
To prove this statement, we here consider only a simple case 
where $f$ is obtained from $g$ by the equation 
$f(x_3,\ldots,x_k) = g^{x_1=c_1,x_2=c_2}(x_3,\ldots,x_k)$, where 
$k\geq3$ and $c_1,c_2\in\{0,1\}$. A more general case can be treated similarly.  
Let $G$ be any undirected bipartite graph that represents $f$ with nodes having labels $x_3,\ldots,x_k$. We construct another bipartite graph $G'$ as follows. We prepare two ``new'' nodes whose labels are $x_1$ and $x_2$. Remember that these variables do not appear in the argument place of $f$. 
Add these new nodes into $G$, replace the node $f$ in $G$ by a ``new'' node labeled $g$ together with two extra edges incident on the nodes $x_1$ and $x_2$, and finally attach two ``new'' nodes with labels $\Delta_{c_1}$ and $\Delta_{c_2}$ to the nodes $x_1$ and $x_2$, respectively, by two ``new'' edges. Clearly, $G'$ realizes $f$ by $\{g,\Delta_{c_1},\Delta_{c_2}\}$. 
Now, let us analyze the node degrees.  Each node $x_i$ ($3\leq i\leq k$) in $G'$ has the same degree as the original node $x_i$ in $G$ does. 
In contrast, the nodes $x_1$ and $x_2$ have only two incident edges.  
Therefore, we conclude that 
$f\leq_{con}^{+0}\{g,\Delta_{c_1},\Delta_{c_2}\}$. 
\end{example}

Unlike the case of T-constructibility, the property of {\em transitivity}  does not hold for limited T-constructibility. Nonetheless, the following restricted form of transitivity is sufficient for our later arguments. 

\begin{lemma}\label{T-transitivity}
Let $f$ and $g$ be any two constraints and let $\GG_1$ and $\GG_2$ be any two finite constraint sets. Moreover, let $d$ be any number in $\nat$. If $f\leq_{con}^{+d}\GG_1\cup\{g\}$ and $g\leq_{con}^{+0}\GG_2$, then $f\leq_{con}^{+d}\GG_1\cup\GG_2$. 
\end{lemma}

\begin{proof}
If $g$ is already in $\GG_1\cup\GG_2$, then the lemma is trivially true; henceforth, we assume that $g\not\in\GG_1\cup\GG_2$. 
Now, let $f(x_1,x_2,\ldots,x_k)$ be any constraint of arity $k\geq1$ and 
let $G_f$ be any undirected bipartite graph, comprised of 
$k$ nodes labeled $x_1,\ldots,x_k$ and a node labeled 
$f$, that represents $f$.  
Assume that $m\geq2$ and each node $x_i$ ($i\in[k]$) 
on the left-hand side of $G_f$ has degree at most $m$. 
Since $f\leq_{con}^{+d}\GG_1\cup\{g\}$, there exists another 
undirected bipartite 
graph $G'_f=(V_1|V_2,E)$ that realizes $f$ by $\GG_1\cup\{g\}$. 
For simplicity, let $V_1=\{x_1,x_2,\ldots,x_k,y_1,y_2,\ldots,y_m\}$ with  
$m$ variables $y_1,\ldots,y_m$ not appearing in $G_f$.  
Note that, by the degree requirement of limited T-constructibility, 
every node $x_i$ ($i\in[k]$) has degree at most $m+d$ and every node $y_j$ ($j\in[m]$) has degree at most $\max\{3,d\}$. 

Since there may be one or more nodes in $G'_f$ whose labels are $g$, we want to eliminate recursively  those nodes one by one. 
Choose any such node, say, $w$. 
We first remove from $G'_f$ all nodes in $V_1\cup V_2$ that are not adjacent to $w$ and also remove their incident edges; however, we keep, as dangling edges, all edges between the remaining nodes in $V_1$ and the 
nodes other than $w$ in $V_2$.  Let $\tilde{G}=(V'_1|V'_2,E')$ be the resulting graph from $G'_f$. Since $V'_1$ is the set of remaining nodes in $V_1$, without loss og generality, we assume that  $V'_1=\{x_1,\ldots,x_a,y_1,\ldots,y_b\}$, where $0\leq a\leq k$ and $0\leq b\leq m$. Since 
$g$ takes all those variables, $\tilde{G}$ obviously represents $g$.  
In this graph $\tilde{G}$, since $f\leq_{con}^{+d}\GG_1\cup\{g\}$, 
every node $x_i$ must have degree at most $m+d$ 
while each node $y_j$ has degree at most $\max\{3,d\}$. 
Since $g\leq_{con}^{+0}\GG_2$, there is another bipartite graph $\tilde{G}'=(V''_1|V''_2,E'')$ that realizes $g$ by $\GG_2$. 
Now, assume that $V''_1=\{x_1,\ldots,x_a,y_1,\ldots,y_b,z_1,\ldots,z_c\}$ with ``fresh'' variables $z_1,\ldots,z_c$. 
 Note that the degrees of the nodes $x_i$ and $y_j$ in $\tilde{G}'$ are the same as that in $\tilde{G}$, and the degree of any other node $z_i$ in $V''_1$ is at most three. Inside $\tilde{G}_f$, we then replace the subgraph $\tilde{G}$ by $\tilde{G}'$. Clearly, the resulting graph has fewer nodes with the label $g$ than $\tilde{G}_f$ does. 
We continue this elimination process until the nodes labeled $g$ are all removed. 
 
In the end, let $G_*$ be the obtained bipartite graph. On the right-hand side of $G_*$, there are only nodes whose labels are taken from $\GG_1\cup\GG_2$.  By its definition, $G_*$ realizes $f$ by $\GG_1\cup\GG_2$. Moreover, in this graph $G_*$, the degree of every node $x_i$ 
is still at most $m+d$ whereas any other node has degree at most $\max\{3,d\}$. 
Therefore, we conclude that $f\leq_{con}^{+d}\GG_1\cup\GG_2$, as requested. 
\end{proof}

\section{Constructing AP-Reductions to the Equality}\label{sec:APreduction-EQ}

Dyer \etalc~\cite{DGJR09} analyzed the complexity of 
approximately solving unweighted 
bounded-degree Boolean \#CSPs and proved  
the first approximation-complexity classification theorem 
for those \#CSPs using 
notions of ``3-simulatability'' and ``ppp-definability.'' 
In their classification theorem, stated in Section \ref{sec:introduction}, they recognized four fundamental categories of counting 
problems. 
We intend to extend their theorem from unweighted 
\#CSPs to complex-weighted \#CSPs by employing the notion of limited T-constructibility described in Section \ref{sec:T-reduction}. 
Our goal is therefore to prove our main theorem, Theorem \ref{main-theorem}. 

We start with a brief discussion on the polynomial-time computability of bounded-degree Boolean \#CSPs. 
For any constraint set $\FF$, it is already known from \cite{Yam10} that, when   $\FF\subseteq\ED$, $\sharpcspstar(\FF)$ is solvable in polynomial time  and thus belongs to $\fp_{\complex}$. {}From this computability result, since $\sharpcspstar_{d}(\FF)\APreduces \sharpcspstar(\FF)$, the following statement is immediate.

\begin{lemma}\label{AF-ED-to-FP}
For any constraint set $\FF$ and any index $d\geq2$, if  $\FF\subseteq\ED$, then $\sharpcspstar_{d}(\FF)$ belongs to $\fp_{\complex}$. 
\end{lemma}

The remaining case where $\FF\nsubseteq \ED$ is the most 
challenging one in this paper. 
In what follows, we are focused on  this difficult case.  
At this point, we are ready to describe an outline of our proof of the main theorem. For notational convenience, we 
write $\EQ$ for the set $\{EQ_k\}_{k\geq2}$, where we do not include the equality of arity $1$, 
because it is in $\UU$ and is always available for free of charge. 
Cai \etalc~\cite{CLX09@} first laid out a basic scheme of how to prove a classification theorem for complex-weighted degree-$3$ Boolean \#CSPs. Later, this scheme was modified by Dyer \etalc~\cite{DGJR09} to prove their classification theorem for unweighted degree-$d$ Boolean \#CSPs for any $d\geq3$.  Our proof strategy closely follows theirs even though we deal with {\em weighted} degree-$d$ \#CSPs.  

For a technical reason, it is better for us to introduce a 
notation $\sharpcspstar_{d}(\EQ\|\FF)$, which is induced from $\sharpcspstar_{d}(\EQ,\FF)$,  
by imposing the following extra 
condition (assuming $\FF\cap\EQ=\setempty$): 
\begin{quote}
(*) In each constraint frame $\Omega=(G,X|\FF',\pi)$ given as input instance instance, no two nodes labeled $EQ$s in $\EQ$ 
(possibly having different arities) 
on the right-hand side of the undirected bipartite graph $G$ are adjacent to the same node having a variable label on the left-hand side of the graph. 
\end{quote}
In other words, any two nodes with labels from $\EQ$ on the right-hand side  of $G$ are not linked directly by any single node. 
This artificial condition (*) is necessary in the proof of Lemma \ref{T-construct-bound}. Similarly, we define 
$\sharpcspstar_{d}(EQ_{k}\|\FF)$ using the singleton $\{EQ_{k}\}$ 
instead of $\EQ$. 
Our proof strategy comprises  
the following four steps. 
\begin{enumerate}\vs{-1}
\item First, for any constraint set $\FF$, we will claim that $\sharpcspstar(\FF)\APequiv \sharpcspstar(\FF')$, where $\FF'=\FF-\EQ$. 
Meanwhile, we will focus on this set $\FF'$. 
Second, we will add the equality of various arity 
and then reduce the original \#CSPs to bounded-degree \#CSPs 
with the above-mentioned condition (*). 
More precisely, we will AP-reduce $\sharpcspstar(\FF')$ to $\sharpcspstar_{2}(\EQ\|\FF')$. 
\vs{-2}
\item For any index $d\geq2$ and for any constraint $f\in\FF$, 
we will AP-reduce 
$\sharpcspstar_{2}(EQ_d\|\FF')$ to $\sharpcspstar_{3}(f,\FF')$, which is clearly AP-reducible to $\sharpcspstar_{3}(\FF)$ since $\{f\}\cup \FF'\subseteq \FF$. 
In addition, we will demand that this reduction should be  
algorithmically ``generic'' 
and ``efficient'' so that if we can AP-reduce $\sharpcspstar_{2}(EQ_d\|\FF')$ to $\sharpcspstar_{3}(f,\FF')$ 
for every index $d\geq3$, then we immediately 
obtain $\sharpcspstar_{2}(\EQ\|\FF')\APreduces \sharpcspstar_{3}(f,\FF')$.  
\vs{-2}
\item Combining the above two AP-reductions, we obtain the AP-reduction  $\sharpcspstar(\FF)\APreduces \sharpcspstar_{3}(\FF)$ by Lemma \ref{AP-properties}.   
Since $\sharpcspstar_{3}(\FF)\APreduces \sharpcspstar_{d}(\FF) 
\APreduces \sharpcspstar(\FF)$ for any index $d\geq3$, 
we conclude that $\sharpcspstar(\FF)\APequiv \sharpcspstar_{d}(\FF)$.  
This becomes our key claim, Proposition \ref{equivalence-EQ}. 
\vs{-2}
\item Finally, we will apply the dichotomy theorem \cite{Yam10} for 
$\sharpcspstar(\FF)$'s to determine the approximation complexity of $\sharpcspstar_{d}(\FF)$'s using the key claim stated in Step $3$. 
\end{enumerate}\vs{-1}

The first step of our proof strategy described above is quite easy 
and we intend to present it here. 

\begin{lemma}\label{F-reduced-to-EQ}
Let $\FF$ be any constraint set and define $\FF'=\FF-\EQ$. 
\begin{enumerate}\vs{-2}
\item $\sharpcspstar(\FF)\APequiv \sharpcspstar(\FF')$.
\vs{-2}
\item $\sharpcspstar(\FF') \APreduces \sharpcspstar_{2}(\EQ\|\FF')$.
\end{enumerate}
\end{lemma}

\begin{proof}
(1) Obviously, it holds that $\sharpcspstar(\FF')\APreduces \sharpcspstar(\FF)$ because $\FF'\subseteq\FF$. What still remains is to  build the opposite AP-reduction. 
Now, let $\Omega$ be any constraint frame given to $\sharpcspstar(\FF)$ with an undirected bipartite graph $G=(V_1|V_2,E)$, where all nodes in $V_1$ have variable labels.  
Note that, whenever there is a node $v$ labeled $EQ_d$ ($d\geq2$) in $V_2$ that has two or more edges incident on the {\em same} node in $V_1$, we can delete all but one such edge without changing 
the outcome of $\csp_{\Omega}$.  To keep the node labeling valid, we need to replace the label $EQ_d$ by $EQ_{d'}$, where $d'$ equals $\deg(v)$ in the modified graph.  In the following argument, we assume that any node with label $EQ_d$ in $V_2$ is always adjacent to $d$ {\em distinct} 
nodes in $V_1$.  

Choose any node, say, $v$ whose label is $EQ_d$ ($d\geq2$) in $V_2$. Let us consider a subgraph $G_v$ consisting only of the node $v$ and of all nodes labeled, say, $x_1,\ldots,x_d$ adjacent to $v$, together with all edges  between $v$ and those $d$ nodes. The graph $G'$ is also composed of, as dangling edges, all edges that have been linked between any node $x_i$ ($i\in[d]$) and any node in $V_2-\{v\}$.   
We first observe that all values of the variables  $x_1,\ldots,x_d$ should coincide in order to make $EQ_d(x_1,\ldots,x_d)$ non-zero. {}From this property, we  merge all the nodes $x_1,\ldots,x_d$ into a single node $w$ with a ``new''  variable label, say, $x'$ and then delete all edges but one that become incident on both $w$ and $v$, while we keep the dangling edges as all distinct edges. Finally, we label the node $v$ by $EQ_1$.  Let $G'_v$ be the graph induced from $G_v$ by the above modification. Now, we replace $G_v$ that appears as a subgraph inside $G$ by $G'_v$. 
This replacement process is repeated until all nodes labeled $EQ_d$ ($d\geq2$) are removed. The obtained graph $G'$ has no node whose label is taken from $\EQ$. Let $\Omega'$ be the constraint frame associated with $G'$.  Since the replacement does not change the value of $\csp_{\Omega}$,  $\csp_{\Omega'} = \csp_{\Omega}$ follows, and thus we obtain $\sharpcspstar(\FF)\APreduces \sharpcspstar(\FF')$.  
 
(2) Given an input instance $\Omega=(G,X|\FF',\pi)$ to  
$\sharpcspstar(\FF')$ with $G=(V_1|V_2,E)$, we will construct another instance $\Omega'$ 
to $\sharpcspstar_{2}(\EQ\|\FF)$ by applying the following recursive procedure. Choose any node of degree $d$ ($d\geq2$) in $V_1$ 
and assume that this node has label $x$.  
Let $e_1,\ldots,e_d$ be the $d$ distinct edges incident on this node $x$ and assume that each $e_i$ ($i\in[d]$) bridges between the node $x$ and a node labeled, say, $g_i$ in $\FF'$.   
Delete this node $x$ and replace it with $d$ ``new'' nodes having 
variable labels, say, $y_1,y_2,\ldots,y_d$ that do not appear in $G$. 
Introduce an additional ``new'' node, say, $v$ labeled $EQ_d$ to $V_2$. 
For each index $i\in[d]$, we re-attach to node $y_i$ 
each edge $e_i$ from the node $g_i$ 
and then make all the nodes $y_1,\ldots,y_d$ adjacent to the node $v$ 
by $d$  ``new'' edges. Notice that 
each node $y_i$ ($i\in[d]$) is now adjacent to two nodes $v$ and $g_i$. 
We continue this procedure 
until all original nodes of degree at least two in $V_1$ are replaced.  

To the end, let $G'$ denote the obtained bipartite graph from $G$ 
and let $\Omega'$ be its associated constraint frame. By our construction, 
any node on the left-hand side of $G'$ has degree exactly two. In addition, 
no two nodes labeled $EQ_d$ share the same variables. 
Since $\csp_{\Omega} = \csp_{\Omega'}$ obviously holds, 
the lemma thus follows. 
\end{proof}

The reader might wonder why we have used $\EQ$, instead of $\{EQ_{2}\}$,  
in the above lemma  although any $EQ_{d}$ 
can be expressed by a finite chain of $EQ_{2}$'s; for instance, $EQ_3(x_1,x_2,x_3)$ equals $EQ_2(x_1,x_2)EQ_2(x_2,x_3)$.  
The reason we have not used $EQ_{2}$ alone 
in (2) of the above proof is that, after running the construction procedure in (2), any node with a variable label that directly connects two $EQ_{2}$'s  
becomes degree three instead of two, and thus this fact proves  
$\sharpcspstar(\FF')\APreduces \sharpcspstar_{3}(EQ_{2}\|\FF')$, 
from which we deduce $\sharpcspstar(\FF)\APequiv \sharpcspstar_{4}(\FF)$.  This consequence is clearly weaker than 
what we wish to establish.  

In the second step of our strategy, we plan to define an AP-reduction from 
$\sharpcspstar_{2}(EQ_d\|\FF)$ to $\sharpcspstar_{3}(\GG,\FF)$. For this purpose, it suffices to prove, as a special case of the following lemma, that $EQ_{d}\leq_{con}^{+1}\GG$ by a generic and efficient algorithm. 
 
\begin{lemma}\label{T-construct-bound}
Let $d,m\in\nat$ with $d\geq2$. Let $\FF$ and $\GG$ be any two constraint 
sets and assume that $\FF\cap\EQ=\setempty$ and $\GG$ is finite. If $EQ_{d}\leq_{con}^{+m}\GG$, then $\sharpcspstar_{2}(EQ_{d}\|\FF)\APreduces \sharpcspstar_{2+m}(\GG,\FF)$. In addition, assume that there exists a procedure of transforming any graph $G$ representing $EQ_d$ into another graph $G'$ realizing $EQ_d$ by $\GG$ in time polynomial in the size of $d$ and the size of the graph $G$. It therefore holds that $\sharpcspstar_{2}(\EQ\|\FF) \APreduces \sharpcspstar_{2+m}(\GG,\FF)$. 
\end{lemma}

\begin{proof}
Let $\Omega$ be any constraint frame given 
as an input instance to $\sharpcspstar_{2}(EQ_{d}\|\FF)$, including 
an undirected bipartite graph $G=(V_1|V_2,E)$.  
Similarly to the proof of Lemma \ref{F-reduced-to-EQ}(1), 
we hereafter assume that any node with label $EQ_d$ in $V_2$ is 
adjacent to $d$ {\em distinct} nodes in $V_1$.  

Now, we will describe a procedure of how to generate a new instance $\tilde{\Omega}$ to $\sharpcspstar_{2+m}(\GG,\FF)$. 
Let $D$ be the collection of all nodes in $V_2$ whose labels are $EQ_d$. 
The following procedure will remove all nodes in $D$ recursively. 
Let us pick an arbitrary node $v$ in $D$ and 
consider any subgraph $G'$ of $G$ satisfying that 
$G'$ consists only of the node $v$ and $d$ different nodes labeled, say, $x_{i_1},\ldots,x_{i_d}$ in $V_1$ that are all adjacent to $v$. 
Because of the degree bound of $\sharpcspstar_{2}(EQ_{d}\|\FF)$, 
each of those $d$ nodes on the left-hand side of $G'$  
should contain at most one dangling edge, which is originally 
incident on a certain other node in $V_2$.  
Clearly, $G'$ represents $EQ_d$. 
Since $EQ_{d}\leq_{con}^{+m}\GG$, there exists another undirected 
bipartite graph $G''$ that realizes $EQ_d$ by $\GG$. 
Inside the original graph $G$, we replace this subgraph $G'$ by $G''$. 
Note that, in this replacement, any node other than 
$x_{i_1},\ldots,x_{i_d}$ in $G''$ are treated as ``new'' nodes; thus, 
those new nodes are not adjacent to any node outside of $G''$.
Furthermore, for each dangling edge appearing in $G'$, we restore its original edge connection to a certain node in $V_2$. Clearly, the resulting graph contains less nodes having the label $EQ_d$. The above process 
is repeated until all nodes in $D$ are removed.  

Let $\tilde{G}$ be the bipartite graph obtained by applying the aforementioned procedure  
and let $\tilde{\Omega}$ be the new constraint frame associated with $\tilde{G}$. The degree of each node $x_{i}$ in $\tilde{G}$ is at most 
$m$ plus the original degree in $G$ since no two nodes labeled $EQ_{d}$ in $G$ share the same variables. By the realizability notion, it is not difficult to show that $\csp_{\tilde{\Omega}} = \csp_{\Omega}$. This implies that $\sharpcspstar_{2}(EQ_d|\FF) \APreduces \sharpcspstar_{2+m}(\GG,\FF)$. 

The second part of the lemma comes from the fact that, using the procedure described above,  we can construct $\tilde{\Omega}$ from $\Omega$ efficiently and robustly if there is a generic procedure that transforms $G'$ to $G''$ for any degree-bound $d$ in polynomial time. Since the premise of the lemma guarantees the existence of such a generic procedure, we immediately obtain the desired consequence. 
\end{proof}

\section{Basic AP-Reductions of Binary Constraints}\label{sec:arity-2}

Since we have shown in Section \ref{sec:APreduction-EQ} that $\sharpcspstar(\FF)$ can be AP-reduced to $\sharpcspstar_{2}(\EQ\|\FF')$, 
where $\FF'=\FF-\EQ$, the remaining task is to  AP-reduce $\sharpcspstar_{2}(\EQ\|\FF')$ further to $\sharpcspstar_{3}(f,\FF')$. To fulfill this purpose, it suffices to prove that, for any index $d\geq2$ and for any constraint $f\in\FF$, $EQ_{d}$ is limited 
T-constructible from $f$ together with (possibly) a few extra unary constraints while 
maintaining the degree-bound to three.  To be more precise, we want to prove that there exists a finite set $\GG\subseteq \UU$ for which  $EQ_{d}\leq_{con}^{+1}\GG\cup\{f\}$. 

By examining the proofs of each lemma given below, it is easy to check 
that the procedure of showing a limited T-constructibility 
relation  $EQ_{d}\leq_{con}^{+1}\GG\cup\{f\}$ for each index 
$d\geq3$ is indeed  ``generic'' 
and ``efficient,'' as requested by Lemma \ref{T-construct-bound}. Therefore, we will finally conclude that 
$\sharpcspstar_{2}(\EQ\|\FF')\APreduces \sharpcspstar_{3}(f,\FF')$. 

This section deals only with non-degenerate constraints of arity two, because degenerate constraints have been already handled by Lemma 
\ref{AF-ED-to-FP}.  
The first case to discuss is a constraint $f$ of the form 
$(0,a,b,0)$ with $ab\neq0$, whose underlying relation $R_f$ is  
$XOR$. 

\begin{lemma}\label{0-a-b-0-case}
Let $d$ be any index at least two. Let $f=(0,a,b,0)$ with $a,b\in\complex$. If $ab\neq0$, then  $EQ_{d}\leq_{con}^{+1}f$ holds.  
\end{lemma}

\begin{proof}
\sloppy {}From a given constraint $f=(0,a,b,0)$, we define another constraint $g$ as $g(x_1,x_2) = \sum_{y\in\{0,1\}}f(x_1,y)f(y,x_2)$. A direct calculation shows that $g= (ab,0,0,ab)$. {}From this definition of $g$, 
we note that (*) the value of $y$ is uniquely determined from $(x_1,x_2)$ if $g(x_1,x_2)\neq0$. 
More generally, for each index $d\geq2$, we define 
$h(x_1,\ldots,x_d) = \sum_{y_1,\ldots,y_{d-1}\in\{0,1\}} \prod_{i=1}^{d-1}(f(x_i,y_i)f(y_i,x_{i+1}))$. 
Clearly, when $d=2$, $h$ coincides with $g$.  
Because of the uniqueness property of $g$ stated in (*), $h(x_1,\ldots,x_d)$ equals $\prod_{i=1}^{d-1}g(x_i,x_{i+1})$. This implies that $h(0,\ldots,0)=h(1,\ldots,1)= (ab)^{d-1}$ and $h(e)=0$ for any other variable assignment $e\in\{0,1\}^d$. It therefore follows that 
$h=(ab)^{d-1}\cdot EQ_{d}$.  Since $ab\neq0$, by normalizing $h$ appropriately, we then obtain $EQ_d$ from $h$.

Next, we will show that $EQ_{d}\leq_{con}^{+1}f$. Let $G$ be any undirected bipartite graph representing $EQ_d$ with $d$ nodes 
whose labels are $x_1,\ldots,x_d$. Consider a new graph $G'$ obtained from $G$, using the above equation of $h$, by adding $d-1$ ``new'' nodes labeled $y_1,\ldots,y_{d-1}$ and by replacing the node $EQ_d$ in $G$ with $2(d-1)$ ``new'' nodes labeled $f$,   
each of which is adjacent to two nodes $x_i$ and $y_i$ ($i\in[d-1]$) or two nodes $y_i$ and $x_{i+1}$.  
This bipartite graph $G'$ clearly realizes $EQ_d$ by $f$. 
Two special nodes $x_1$ and $x_d$ in $G'$ maintain their original degree in $G$, whereas each node $x_i$ except for $x_1$ and $x_d$ has one more than its original degree in $G$. 
In addition, all nodes with the labels $y_1,\ldots,y_{d-1}$ are of  
degree exactly two. Therefore, we conclude that 
$EQ_{d}\leq_{con}^{+1}f$, as requested.  
\end{proof}

As the second case, we will handle a constraint $f=(a,0,0,b)$ satisfying  $ab\neq0$. Since its underlying relation is precisely $EQ_2$,  the proof of 
its limited T-constructibility is rather simple. 

\begin{lemma}\label{a-0-0-b-case}
Let $d\geq2$ and let $f=(a,0,0,b)$ with $a,b\in\complex$. If $ab\neq0$, then there exists a constraint $u\in\UU\cap\NZ$ such that $EQ_d\leq_{con}^{+1}\{f,u\}$. 
\end{lemma}

\begin{proof}
Let $f=(a,0,0,b)$ with $ab\neq0$. First, we consider the base case of 
$d=2$. By setting $u=[1/a,1/b]$, we define a constraint $g$ as 
$g(x_1,x_2) = u(x_1)f(x_1,x_2)$. 
Clearly, $g$ equals $EQ_2$. For a degree analysis, 
let us consider any undirected bipartite graph $G$ that represents $EQ_d$. Since $g= EQ_2$,  a new bipartite graph $G'$ is obtained from $G$ 
by replacing the existing node $EQ_d$ and its associated edges in $G$ with 
two ``new'' nodes labeled $u$ and $f$ together with three ``new'' edges $\{(x_1,u),(x_1,f),(x_2,f)\}$. 
{}From this construction, the node $x_1$ in $G'$ has one 
more than its original degree in $G$; however, the degree of the node $x_2$ in $G'$ remains the same as that in $G$. We therefore 
obtain $EQ_2\leq_{con}^{+1}\{f,u\}$. This argument will be extended  to the general case of $d\geq 2$.  

For each fixed index $d\geq2$, we set $u'=[1/a^d,1/b^d]$ and define 
$h(x_1,\ldots,x_d) = u'(x_1) \prod_{i=1}^{d-1} f(x_i,x_{i+1})$. 
It is not difficult to show that $h$ equals $EQ_{d}$. 
Similarly to the base case, from the definition of $h$, we can build 
a bipartite graph $G'$ that realizes $h$ by $\{f,u'\}$. 
In this graph $G'$, each node $x_i$ ($1\leq i<d$) has one more than its original degree in $G$, while the node $x_d$ keeps the same degree  as that in $G$. This fact helps us conclude that  $EQ_{d}\leq_{con}^{+1}\{f,u'\}$. 
\end{proof}

Our next target is a constraint $f$ of the form $(a,b,0,c)$ with $abc\neq0$. The underlying relation of $f$ is exactly $Implies$. 

\begin{lemma}\label{a-b-0-c-case}
Let $d\geq2$. Let $f=(a,b,0,c)$ with $a,b,c\in\complex$. 
If $abc\neq0$, then  there exist two constraints $u_1,u_2\in\UU\cap\NZ$ 
for which $EQ_{d}\leq_{con}^{+1}\{f,u_1,u_2\}$. 
By permuting variable indices, the case of $(a,0,b,c)$ is similar. 
\end{lemma}

\begin{proof}
First, we set $f=(a,b,0,c)$ and assume that $abc\neq0$. 
For this constraint $f$, we prepare the following 
two unary constraints: $u=[1/a^2,1/c^2]$ and 
 $u'=[1/a^3,1/c^3]$. Let us begin with the base case of $d=2$. 
In this case, we define $g(x_1,x_2) = f(x_2,x_1)\sum_{y\in\{0,1\}} f(x_1,y)u'(y)f(y,x_2)$. Since $u'$ 
cancels out the effect of both terms $f(x_2,x_1)$ and 
$f(x_1,y)f(y,x_2)$, we immediately obtain $g=(1,0,0,1)$. 

Let $G$ be any undirected bipartite graph representing $EQ_2$ with two variables $x_1$ and $x_2$.  
To obtain another bipartite graph $G'$ realizing $EQ_2$, 
we first build a graph $G^{[f,u']}$ (using $u'$ instead of $u$),  
introduced in Example \ref{realize-g}, which is equipped with all the 
original dangling edges in $G$. We next add 
an extra ``new'' node with label $f$ that becomes adjacent 
to the two nodes $x_2$ and $x_1$.   
This newly constructed graph $G'$ obviously realizes $EQ_2$ by  
$\{f,u'\}$. Since $G'$ contains two edges from each node $x_i$ ($i\in\{[2]$), 
the degree of the node $x_i$ in $G'$ thus increases by one, and therefore  $EQ_2\leq_{con}^{+1}\{f,u'\}$ follows. 

In the case of $d\geq3$, by extending the base case, 
we naturally define a constraint $h$ as $h(x_1,\ldots,x_d) = f(x_d,x_1) \sum_{y_1,\ldots,y_{d-1}\in\{0,1\}} \prod_{i=1}^{d-1}(f(x_i,y_i)u_i(y_i)f(y_i,x_{i+1}))$, where $u_{d-1}= u'$ and 
$u_i= u$ for each $i\in[d-2]$. Note that $u$ and $u'$ bring the same effect as $u'$ does in the base case. 
The analysis of the node degrees in the corresponding graph is similar 
in essence to the degree analysis of the base case. Therefore, it immediately follows that $EQ_d\leq_{con}^{+1}\{f,u,u'\}$. 
\end{proof}

Unlike the constraints we have discussed so far, 
the non-degenerate non-zero constraints $f=(1,a,b,c)$ 
with $a,b,c\in\complex$   
are quite special, because they appear only in the case of 
complex-weighted \#CSPs. When $f$ is limited to be a Boolean relation, 
by contrast, it never becomes both non-degenerate and non-zero. 
Notice that, by Lemma \ref{basic-DG-IMP}, $f\not\in\DG$ is  equivalent to $ab\neq c$.

\begin{lemma}\label{1-a-b-c-case}
Let $d\geq2$ and let $f=(1,a,b,c)$ with $abc\neq0$. If $ab\neq c$, 
then there exist two constraints $u_1,u_2\in\UU\cap\NZ$ satisfying 
that $EQ_{d}\leq_{con}^{+1}\{f,u_1,u_2\}$. 
\end{lemma}

\begin{proof}
Let $f=(1,a,b,c)$ be any binary constraint satisfying that $abc\neq0$ and $ab\neq c$.  Now, we set $u_1=[1,z]$ and define $g$ as $g(x_1,x_2) = \sum_{y\in\{0,1\}} f(x_1,y)u_1(y)f(y,x_2)$. 
This gives $g=(1+abz,a(1+cz),b(1+cz),ab+c^2z)$. 
If we choose $z=-1/c$, then the constraint   
$g$ becomes of the form $(1-ab/c,0,0,ab-c)$. 
Note that, since $ab\neq c$, the first and last entries of $g$ are non-zero. 
By appealing to (the proof of) Lemma \ref{a-0-0-b-case}, which requires another 
non-zero unary constraint $u_2$, the new constraint $g'(x_1,x_2) = u_2(x_1)g(x_1,x_2)$ equals $EQ_2(x_1,x_2)$. 

To show $EQ_{2}\leq_{con}^{+1}\{f,u_1,u_2\}$, from any undirected bipartite graph $G$ representing $EQ_2$ with variables $x_1$ and $x_2$, we construct another graph $G'$ by taking $G^{[f,u_1]}$ (stated in Example \ref{realize-g}) with the original dangling edges in $G$ and 
further by adding a ``new'' node labeled $u_2$ that is adjacent to the node $x_1$. Overall, the degree of any node on the left-hand side of $G'$ 
increases by at most one in 
comparison with the degree of the same node in $G$. 

In a more general case of $d\geq3$, with a series $x=(x_1,\ldots,x_d)$ 
of $d$ variables, we define $g(x) = 
\sum_{y_1,\ldots,y_{d-1}\in\{0,1\}} \prod_{i=1}^{d-1}(f(x_i,y_i)u_1(y_i)f(y_i,x_{i+1}))$. Since $g$ has the form $(a',0,\ldots,0,b')$,  with an appropriate constraint 
$u'_2\in\UU\cap\NZ$, the constraint $g'(x) = u'_2(x_1)g(x)$ coincides with  $EQ_d$.  
A degree analysis of a graph realizing $EQ_d$ is similar to the base case. We therefore obtain $EQ_d\leq_{con}^{+1}\{f,u_1,u'_2\}$.   
\end{proof}

As a summary of Lemmas \ref{0-a-b-0-case}--\ref{1-a-b-c-case}, we wish to make a general claim on 
binary constraints that do not belong to $\DISJ\cup\NAND\cup\DG$.  This claim will be a basis of the proof of Proposition \ref{outside-DISJ-NAND}. 

\begin{proposition}\label{basis-case-not-DISJ}
Let $d\geq2$. For any non-degenerate  binary 
constraint $f$, if $f\not\in \DISJ\cup\NAND\cup\DG$, then there exists a constraint set $\GG\subseteq \UU\cap\NZ$ with $|\GG|\leq2$ such that $EQ_{d}\leq_{con}^{+1} \GG\cup\{f\}$.     
\end{proposition}

\begin{proof}
Let $f=(a,b,c,d)$ be any non-degenerate constraint.  It is important to note that $f\not\in\DISJ\cup\NAND$ 
iff $f$ is one of the following forms: $(0,b,c,0)$, $(a,0,0,d)$, $(a,0,c,d)$, $(a,b,0,d)$, and $(a,b,c,d)$, 
provided that $abcd\neq0$.  In particular, for the last form $(a,b,c,d)$, 
since $f\not\in\DG$, Lemma \ref{basic-DG-IMP} yields the inequality $ad\neq bc$. All the above five forms have been already 
dealt with in Lemmas \ref{0-a-b-0-case}--\ref{1-a-b-c-case}, and therefore the lemma should hold. 
\end{proof}

The most notable case is where $f=(0,a,b,c)$ or $f=(a,b,c,0)$ with $abc\neq0$. These two constraints respectively extend $OR_2$ and $NAND_2$ from Boolean values to complex values. 
Our result below contrasts complex-weighted constraints with unweighted constraints, because this result is not known to hold for the Boolean constraints. 

\begin{proposition}\label{OR-is-enough}
Let $d\geq2$. If $f= (0,a,b,c)$ with $abc\neq0$, then there exists a constraint $u\in\UU\cap\NZ$ such that $EQ_{d}\leq_{con}^{+1}\{f,u\}$. A similar statement holds for $f= (a,b,c,0)$ with $abc\neq0$. 
\end{proposition}

The proof of this proposition utilizes two useful lemmas, Lemmas \ref{0-a-b-c-case} and \ref{reduce-OR-to-NAND}, which are described below.  
In the first lemma, we want to show that two constraints whose underlying relations are $OR_2$ and $NAND_2$ together help compute $EQ_d$ for any index $d\geq2$.  

\begin{lemma}\label{0-a-b-c-case}
Let $d\geq2$. Let $f_1=(0,a,b,c)$ and $f_2=(a',b',c',0)$ with 
$a,b,c,a',b',c'\in\complex$. If $ab\neq 0$ and $b'c'\neq 0$, then 
$EQ_{d}\leq_{con}^{+1}\{f_1,f_2\}$.  
\end{lemma}

\begin{proof}
Let $f_1=(0,a,b,c)$ and $f_2=(a',b',c',0)$  with $abb'c'\neq0$. 
First, we explain our construction for the base case of $d=2$. 
By defining $g(x_1,x_2) = \sum_{y_1,y_2\in\{0,1\}} f_1(x_1,y_1)f_1(y_2,x_2) f_2(y_1,x_2)f_2(x_1,y_2)$, $g$ becomes of 
the form $(abb'c',0,0,abb'c')$, from which we immediately obtain $EQ_{2}=(1,0,0,1)$ by normalizing it since $abb'c'\neq0$.   
Let $G=(V_1|V_2,E)$ be any undirected bipartite graph representing $EQ_2$. 
Based on the definition of $g$, we will construct an appropriate bipartite graph  $G'$ from $G$ as follows. We first introduce two additional nodes labeled $y_1$ and $y_2$ into $V_1$. In place of the node labeled $EQ_2$ in $V_2$, 
we next add two ``fresh'' nodes with the same label $f_1$, which  respectively become adjacent to the two nodes $x_1$ and $y_1$ and to the two nodes $y_2$ and $x_2$, and we also add two ``fresh'' nodes having the same label $f_2$, which are respectively adjacent to the nodes $y_1$ and $x_2$ and to the  nodes $x_1$ and $y_2$. 
The degree of each node $x_i$ ($i\in[2]$) in 
$G'$ increases by one from its original degree in $G$,  
because each node $x_i$ is linked in $G'$ to the two nodes with labels  
$f_1$ and $f_2$.  Moreover, the new nodes $y_1$ and $y_2$ have degree exactly two. It therefore holds that $EQ_{2}\leq_{con}^{+1}\{f_1,f_2\}$.

In what follows, we assume $d\geq3$ and focus on the case where $d$ is even. 
We will extend the argument used in the base case.  
Let $x=(x_1,\ldots,x_d)$ and $y=(y_1,\ldots,y_d)$ be two series of distinct variables.  We then introduce  two useful constraints 
$g_1$ and $g_2$ defined by $g_1(x,y) =  \prod_{i=0}^{d/2-1}(f_1(x_{2i+1},y_{2i+1})f_1(y_{2i+2},x_{2i+2}))$ and 
$g_2(x,y) =  
\left(\prod_{i=0}^{d/2-1}f_2(y_{2i+1},x_{2i+2})\right) 
\left(\prod_{i=0}^{d/2-2}f_2(x_{2i+3},y_{2i+2})\right)$. 
With these new constraints, we define 
$h(x) =   \sum_{y_1,\ldots,y_{d}\in\{0,1\}} 
g_1(x,y) g_2(x,y)f_2(x_1,y_d)$. 
By a straightforward calculation, it is not difficult to 
check that $h$ truly computes 
$\lambda \cdot EQ_d$ for a certain constant $\lambda\in\complex-\{0\}$.  Similar to the construction of the base case, from a graph $G$ representing $EQ_d$, we can construct a new bipartite graph $G'$ that realizes $EQ_d$ by $\{f_1,f_2\}$. 
The degree of every node $x_i$ ($i\in[d]$) in $G'$ is one more than its original degree in $G$, whereas  
all nodes $y_j$ ($j\in[d]$) in $G'$ are of degree two.  
Thus, we conclude that 
$EQ_{d}\leq_{con}^{+1}\{f_1,f_2\}$. 

When $d$ is odd, we initially introduce a fresh  
variable called $x_{d+1}$ as a ``dummy.''  
After defining $h(x_1,\ldots,x_{d+1})$ as done before, we need to define $h' = h^{x_{d+1}=*}$, which turns out to equal $\lambda'\cdot EQ_d$ for an appropriate non-zero constant $\lambda'$. The degree analysis of $G'$ is similar to the even case. Therefore, the proof is completed. 
\end{proof}

The second lemma ensures that, with a help of unary constraint, 
we can transform a constraint in $\DISJ$ into another in $\NAND$ without  increasing the degree of its realizing graph. This is a special phenomenon not seen for Boolean constraints and it clearly exemplifies a power of the {\em weighted}  
unary constraints.

\begin{lemma}\label{reduce-OR-to-NAND}
For any binary constraint $h\in\NZ$,  there exist 
a binary constraint $h'\in\NZ$ 
and a unary constraint $u\in\NZ$ such that $NAND_{2}\cdot h'\leq_{con}^{+0}\{OR_{2}\cdot h, u\}$. A similar statement holds if we exchange the roles of $OR_{2}$ and $NAND_{2}$. 
\end{lemma}

\begin{proof}
Let $f = OR_{2}\cdot h$ for a given  constraint 
 $h\in\NZ$ of arity two. By normalizing $f$ appropriately, 
we assume, without loss of generality, that $f$ is of the from $(0,a,b,1)$, where $ab\neq0$.   
With a use of an extra constraint $u=[1,z]$, let us define $g(x_1,x_2) = \sum_{y\in\{0,1\}} f(x_1,y)u(y)f(y,x_2)$, which implies 
$g=(abz,az,bz,ab+z)$. 
Hence, if we set $z= -ab$, then $g$ equals $(-(ab)^2,-a^2b,-ab^2,0)$. 
We then define the desired $h'$ as $(-(ab)^2,-a^2b,-ab^2,1)$, which is obviously a non-zero constraint. 
Obviously, $g(x_1,x_2)$ coincides with $NAND_{2}(x_1,x_2) h'(x_1,x_2)$; thus, we obtain $g=NAND_2\cdot h'$.  
  
Next, we want to show that $g\leq_{con}^{+0}\{f,u\}$. 
Against any graph $G$ representing $g$, we define $G'$ to be  
the graph $G^{[f,u]}$, stated in Example \ref{realize-g},  
together with all dangling edges appearing in $G$.  Recall that $G^{f,u]}$ is comprised of nodes labeled $x_1$, $x_2$, and $y$.  
The degree of the node $y$ in $G'$ is three and the other variable nodes have the same degree as their original ones in $G$. 
It therefore follows that $g\leq_{con}^{+0}\{f,u\}$. 
\end{proof}

Finally, we are ready to give the proof of Proposition \ref{OR-is-enough}.

\begin{proofof}{Proposition \ref{OR-is-enough}}
Let $d\geq2$ and let $f=(0,a,b,c)$ with $abc\neq0$. By setting $h=(1,a,b,c)\in\NZ$, we obtain $f(x_1,x_2) = OR_{2}(x_1,x_2)h(x_1,x_2)$. 
By Lemma \ref{reduce-OR-to-NAND}, there are two constraints $u\in\UU\cap\NZ$ and $h'\in\NZ$ of arity two for which $g\leq_{con}^{+0}\{f,u\}$ and 
$g = NAND_{2}\cdot h'$. 
Note that, since $h'\in\NZ$, $g$ should have the form $(a',b',c',0)$ 
for certain constants $a',b',c'\in\complex$ with $a'b'c'\neq0$. 
Now, we apply Lemma \ref{0-a-b-c-case} 
to $f$ and $g$ and then obtain $EQ_{d}\leq_{con}^{+1}\{f,g\}$. 
Combining this with $g\leq_{con}^{+0}\{f,u\}$, Lemma \ref{T-transitivity} draws the desired conclusion that $EQ_{d}\leq_{con}^{+1}\{f,u\}$.  
\end{proofof}

\section{Constraints of Higher Arity}

We have shown in Section \ref{sec:arity-2} that the equality 
$EQ$ of arbitrary arity can be limited T-constructible from 
non-degenerate binary constraints. Here, we want to prove a similar result for constraints of three or higher arities. Since constraints in $\ED$ already fall into $\fp_{\complex}$, it suffices for us 
to concentrate on the following two types of constraints: (i) constraints within $\DISJ\cup\NAND - \DG$ and (ii) constraints outside of $\DISJ\cup\NAND\cup\DG$.  These types will be discussed in two separate subsections. 

\subsection{Constraints in $\DISJ\cup\NAND - \DG$}

First, we will focus our attention on constraints residing in 
$\DISJ\cup\NAND - \DG$.  Proposition \ref{OR-is-enough} has already 
handled binary  
constraints  chosen from  $\DISJ\cup \NAND-\DG$ with an argument that looks  quite different from the unweighted case of Dyer \etalc~\cite{DGJR09}. 
We will show that this result can be extended to constraints of arbitrary high arity. 

\begin{proposition}\label{DISJ-NAND-to-EQ}
Let $k\geq2$ and $d\geq2$. Let $f$ be any $k$-ary 
constraint in $\DISJ\cup\NAND$. 
If $f\not\in\DG$, then there exists a non-zero unary constraint $u$ such that $EQ_{d}\leq_{con}^{+1}\{f,u,\Delta_0,\Delta_1\}$. Moreover, it holds that $\sharpcspstar_{2}(\EQ\|\FF)\APreduces \sharpcspstar_{3}(f,\FF)$ for any constraint set $\FF$ satisfying $\FF\cap\EQ=\setempty$. 
\end{proposition}

Before proving this proposition, we will 
show below a useful lemma, which requires the following terminology. 
The {\em width} of a constraint $f$ in $\DISJ$ ($\NAND$, resp.) is the maximal arity of any factor that appears in a unique or-distinctive 
(nand-distinctive, resp.) factor list for the underlying relation $R_f$.  
For each index $w\geq2$, we denote $\DISJ_{w}$ ($\NAND_{w}$, resp.) the set of all constraints in $\DISJ$ ($\NAND$, resp.) of width {\em exactly} 
$w$. Note that $\DISJ = \bigcup_{w\geq2}\DISJ_{w}$ and $\NAND = \bigcup_{w\geq2}\NAND_{w}$.

\begin{lemma}\label{OR-NAND-reduction}
Let $w\geq2$ be any width index. For any constraint  $f\in \DISJ_{w}$ 
($\NAND_{w}$, resp.), there exists a non-zero 
constraint $h$ of arity $w$ satisfying that $OR_{w}\cdot h\leq_{con}^{+0}\{f,\Delta_1\}$ 
($NAND_{w}\cdot h\leq_{con}^{+0}\{f,\Delta_0\}$, resp.). 
\end{lemma}

\begin{proof}
In this proof, we will show the lemma only for $\DISJ_{w}$ because the other case, $\NAND_w$, is similar. Assume that $w\geq2$.  Let $k\geq2$ and let  $f\in\DISJ_{w}$ be any arity-$k$ constraint with $k$ variables $x_1,\ldots,x_k$.  Notice that the arity 
of $f$ should be more than or equal to $w$. 
We can express $f$ as $R_f\cdot h$ using an appropriate $k$-ary constraint $h\in \NZ$. 
Hereafter, we look into the underlying relation $R_f$. Let us consider a unique or-distinctive factor list $L$ for $R_{f}$. 
Since $L$ should contain 
at least one $OR$ of arity $w$, $f$ does not belong to $\DG$. 
By pinning $f$, we want to construct a constraint $g$ whose underlying relation equals a factor $OR_w$ in $L$. 
For this purpose, we describe below a two-step procedure of 
how to build such a constraint $g$.  

(1) If there exists a factor of the form $\Delta_c(x)$ ($c\in\{0,1\}$) 
in $L$, then, by assigning the value $c$ to the variable $x$, we obtain a pinned constraint $g'=f^{x=c}$. Since the or-distinctiveness forbids both factors $\Delta_c$ and $OR$ in $L$ to share the same variables, this pinning operation makes $g'$ becomes neither an all-$0$ function nor an all-$1$ function. 

(2) After recursively applying (1), we now assume that 
there is no factor of the from 
$\Delta_c$ in $L$.  
Let us choose an $OR_w$ in $L$. For simplicity, by permuting variable indices, we assume that this $OR_{w}$ takes $w$ distinct 
variables $x_1,x_2,\ldots,x_w$. 
By assigning $1$ to all the other variables $x_{w+1},\ldots,x_k$,   
we obtain $g=f^{x_{w+1}=1,\ldots,x_{k}=1}$, which obviously implies $R_g = R_f^{x_{w+1}=1,\ldots,x_k=1}$.  
By Example \ref{pinning-T}, it holds that $g\leq_{con}^{+0}\{f,\Delta_1\}$. 
Since no variable set of any other $OR$ in $L$ becomes a subset of $\{x_1,\ldots,x_w\}$, $R_g$ actually coincides with the given $OR_w$. 

To end the proof, we set $h'=h^{x_{w+1}=1,\ldots,x_{k}=1}$, implying that $h'$ is of arity $w$. With this $h'$, 
the constraint $g$ can be expressed as 
$g = R_g\cdot h'$, and thus $g$ equals $OR_{w}\cdot h'$ since $R_g=OR_{w}$.  Notice that $h'\in\NZ$ since $h\in\NZ$. Moreover, since $g\leq_{con}^{+0}\{f,\Delta_1\}$, the constraint $OR_{w}\cdot h'$ is limited T-constructible from $\{f,\Delta_1\}$. 
This completes the proof of the lemma. 
\end{proof}

Proposition \ref{DISJ-NAND-to-EQ} follows directly from 
Lemma \ref{OR-NAND-reduction} together with Proposition \ref{OR-is-enough}. 

\begin{proofof}{Proposition \ref{DISJ-NAND-to-EQ}}
Assume that $f\in\DISJ$ and $f$ has arity $k$. 
In addition, we assume that $f$ has width $w$ for a certain number  $w\geq2$; namely, $f\in\DISJ_{w}$. Notice that $k\geq w$.  
Lemma \ref{OR-NAND-reduction} ensures the existence of a constraint $h\in\NZ$ of arity $w$ for which $OR_{w}\cdot h\leq_{con}^{+0}\{f,\Delta_1\}$. 

Assume that this relation $OR_w$ takes $w$ distinct variables, say, 
$x_1,\ldots,x_w$. 
We then choose two specific variables,  $x_1$ and $x_2$, 
and assign $0$ to all the other variables. 
Let $f'$ be the constraint obtained from $OR_{w}\cdot h$ 
by performing these pinning operations. By the construction of $f'$, Example \ref{pinning-T} implies $f'\leq_{con}^{+0}\{f,\Delta_0,\Delta_1\}$. 
It is not difficult to show that, since $h\in\NZ$, 
$R_{f'}(x_1,x_2)$ equals $OR_2(x_1,x_2)$; 
in other words, $f'$ is of the form $(0,a,b,c)$ with $abc\neq0$. 
 
Finally, we apply Proposition \ref{OR-is-enough} and 
then obtain a constraint $u\in\UU\cap\NZ$ satisfying that $EQ_{d}\leq_{con}^{+1}\{f',u\}$. 
We combine this with $f'\leq_{con}^{+0}\{f,\Delta_0,\Delta_1\}$ 
to conclude by Lemma \ref{T-transitivity} that   $EQ_{d}\leq_{con}^{+1}\{f,u,\Delta_0,\Delta_1\}$. 
The case where $f\in\NAND$ is similarly treated. 

The second part of the proposition follows by Lemma \ref{T-construct-bound}  from the fact that the above procedure is indeed generic and efficient. 
\end{proofof}

\subsection{Constraints Outside of $\DISJ\cup\NAND\cup\DG$}

The remaining type of constraints to discuss is ones that sit 
outside of $\DISJ\cup\NAND\cup\DG$. As a key claim for those constraints, 
we will prove the following proposition.

\begin{proposition}\label{outside-DISJ-NAND}
Let $d$  and $k$ be any two indices at least two. For any 
constraint $f$ of arity $k$,  
if $f\not\in \DISJ\cup\NAND\cup\DG$, then  
there exists a finite subset $\GG$ of $\UU$ such that 
$EQ_{d}\leq_{con}^{+1}\GG\cup\{f\}$. In addition, it holds that 
$\sharpcspstar_{2}(\EQ \|\FF)\APreduces \sharpcspstar_{3}(f,\FF)$ 
for any constraint set $\FF$ satisfying $\FF\cap\EQ=\setempty$.    
\end{proposition}


This proposition will be proven by induction on the arity of a given constraint $f$.  As our starting point, we want to prove a useful lemma regarding non-degenerate constraints of particular form. 

\begin{lemma}\label{DG-arity-reduction}
Let $k\geq3$. Let $f$ be any non-degenerate constraint of arity $k$. 
If $f^{x_1=0},f^{x_1=1}\in\DG$, then there exists a non-degenerate constraint $h$ of arity $k-1$ for which $h\leq_{con}^{+0}\GG\cup\{f\}$ for a certain finite subset $\GG$ of $\UU\cap\NZ$.  
\end{lemma}

\begin{proof}
For any fixed index $k\geq3$, let us choose any arity-$k$ constraint $f$ not in $\DG$ and set $g_b=f^{x_1=b}$ for every index $b\in\{0,1\}$. Assume that $g_0$ and $g_1$ are degenerate.  First, we define a ``factor list'' for $g_b$.  Since $g_b\in\DG$,  $g_b(x_2,x_3,\ldots,x_k)$ can be expressed as  $\alpha' g_{b,2}(x_2)g_{b,3}(x_3)\cdots g_{b,k}(x_k)$, where  $\alpha'$ is an appropriate constant in $\complex-\{0\}$ and each $g_{b,i}$ has one of the following forms: $\Delta_0(x_i)$, $\Delta_1(x_i)$, and $[1,a](x_i)$ with $a\neq0$.  We call the set  $L_b=\{g_{b,2}(x_2),g_{b,3}(x_3),\ldots,g_{b,k}(x_k)\}$ (ignoring the global constant $\alpha'$) a {\em factor list} for $g_b$.  Such a factor list is obviously unique. 

(1) If $L_0$ and $L_1$ share the same factor of the form, $\Delta_0(x_i)$, $\Delta_1(x_i)$, or $[1,1](x_i)$ for a certain index $i$ with 
$2\leq i\leq k$, then we define $h=f^{x_i=*}$. In case of $\Delta_0(x_i)$, for example, it holds that $f(x_1,x_2,\ldots,x_{k}) = \Delta_0(x_i) h(x_1,\ldots,x_{i-1},x_{i+1},\ldots,x_k)$. {}From this equation, if $h$ is degenerate, then $f$ should be degenerate, contradicting our assumption.  Thus, $h$ cannot be degenerate. The other cases are similar. Obviously, the arity of $h$ is exactly $k-1$. Since $h\leq_{con}^{+0}f$, we immediately obtain the lemma. 

(2) Hereafter, we assume that Case (1) never occurs; namely, $L_0\cap L_1=\setempty$. Let us discuss several cases separately.

(i) Assume that, for a certain index $i$, $L_1$ contains a factor $\Delta_0(x_i)$ and $L_2$ contains $\Delta_1(x_i)$. For ease of the description below, we set $i=2$. By the definition of $g_0$, there exists a degenerate constraint $g'_0$ such that  $g_0(x_2,x_3,\ldots,x_k)$ equals  $\Delta_0(x_2)g'_0(x_3,\ldots,x_k)$. Similarly, $g_0(x_2,x_3,\ldots,x_k)$ is of the form $\Delta_1(x_2)g'_1(x_3,\ldots,x_k)$ for a certain $g'_1\in\DG$. For the desired $h$, we define $h=f^{x_2=*}$, which implies that $h^{x_1=0} = g'_0$ and $h^{x_1=1}=g'_1$. Obviously, $g\leq_{con}^{+0}f$ holds. 
Now, we want to claim that $h\not\in \DG$. Toward a contradiction, we assume otherwise. This yields an equation $h^{x_1=0} = \gamma  \cdot h^{x_1=1}$ for a certain non-zero constant $\gamma$; in other words, $g'_0= \gamma \cdot g'_1$ holds. Let us consider two factor lists $L'_0$ and $L'_1$ for $g'_0$ and $g'_1$, respectively. Since $g'_0= \gamma \cdot g'_1$, those two factor lists must coincide. Since $L'_0\subseteq L_0$ and $L'_1\subseteq L_1$, we conclude that $L_0\cap L_1\neq\setempty$. This is a contradiction against $L_0\cap L_1=\setempty$. Therefore, $h\not\in\DG$ follows. 
This $h$ satisfies the lemma since $h$'s arity is $k-1$. 

(ii) Consider the case where  $L_1$ contains $\Delta_0(x_i)$ and $L_2$ contains $[1,a](x_i)$. As before, we set $i=2$. Assume that  $g_0(x_2,x_3,\ldots,x_k) = \Delta_0(x_2)g'_0(x_3,\ldots,x_k)$ and $g'_1(x_2,x_3,\ldots,x_k) = [1,a](x_2)g'_1(x_3,\ldots,x_k)$ for two degenerate constraints $g'_0$ and $g'_1$. First, we select a non-zero constant $\xi$ for which $1+ a\xi\neq0$. With this constant, we then define $h(x_1,x_3,\ldots,x_k) = \sum_{y\in\{0,1\}} f(x_1,y,x_3,\ldots,x_k)[1,\xi](y)$. A simple calculation shows that $h^{x_1=0} = g'_0$ and $h^{x_1=1} = (1+a\xi)\cdot g'_1$.  
Note that $[1,\xi]\in\UU\cap \NZ$ and    $h\leq_{con}^{+0}\{f,[1,\xi]\}$. 
If $h\in\DG$, then an argument similar to (i) proves that $L_0\cap L_1\neq\setempty$, a contradiction. Hence, we conclude that $h\not\in\DG$, ensuring the lemma. 

(iii) Let us assume that $L_1$ contains $[1,a](x_i)$ and $L_2$ contains $[1,b](x_i)$ with $ab\neq0$. Set $i=2$ for simplicity.  Assume that $g_0$ and $g_1$ are of the form: $g_0(x_2,x_3,\ldots,x_k) = [1,a](x_2)g'_0(x_3,\ldots,x_k)$ and $g_1(x_2,x_3,\ldots,x_k) = [1,b](x_2)g'_1(x_3,\ldots,x_k)$ for certain constraints $g'_0,g'_1\in\DG$. 
To obtain the lemma, here we first choose a non-zero constant $\xi$ 
to satisfy that $\xi+a\neq0$ and $\xi+b\neq0$. 
The desired $h$ is now defined as $h(x_1,x_3,\ldots,x_k) = \sum_{y\in\{0,1\}} f(x_1,y,x_3,\ldots,x_k)[\xi,1](y)$. 
It then holds that $h^{x_1=0} = (\xi+a)\cdot g'_0$ and $h^{x_1=1} = (\xi+b)\cdot g'_1$.  When $h\in\DG$, $(\xi+a)\cdot g'_0 = \gamma(\xi+b)\cdot g'_1$ holds for a non-zero constant $\gamma$. Since both values $\xi+a$ and $\gamma(\xi+b)$ are not zero, a similar argument to (i) leads to a contradiction. Therefore, we obtain  $h\not\in\DG$, as required. 

(iv) The other cases are similar to (i)--(iii).   
\end{proof}

The second step for the proof of 
Proposition \ref{outside-DISJ-NAND} is made by the following lemma.  

\begin{lemma}\label{induction-step-not-DISJ}
Let $d\geq2$ and $k\geq3$. For any $k$-ary constraint 
$f\not\in \DISJ\cup\NAND\cup\DG$, if $EQ_{d}\not\leq_{con}^{+1}\GG\cup\{f\}$ for any finite set $\GG\subseteq\UU$, then there exists another constraint $g$ of arity $k-1$ such that $g\not\in \DISJ\cup\NAND\cup\DG$ and $g\leq_{con}^{+0}\GG'\cup\{f\}$ for a certain finite set $\GG'\subseteq\UU\cap\NZ$.     
\end{lemma}

\begin{proof}
Let $f\not\in \DISJ\cup\NAND\cup\DG$ be any $k$-ary constraint. 
Assume that $EQ_d\not\leq_{con}^{+1}\GG\cup\{f\}$ for any finite set $\GG\subseteq \UU$. With constraints $g_b=f^{x_1=b}$ for two values   $b\in\{0,1\}$, it holds that  $f(x_1,x_2,\ldots,x_k) = \sum_{b\in\{0,1\}}\Delta_{b}(x_1)g_b(x_2,\ldots,x_k)$. 
Obviously, both $g_0$ and $g_1$ have arity $k-1$  and 
$g_b\leq_{con}^{+0}\{f,\Delta_b\}$ holds by Example \ref{pinning-T} 
for any $b\in\{0,1\}$. 
If either $g_0$ or $g_1$ stays out of $\DISJ\cup\NAND\cup\DG$, then we immediately obtain the lemma. Henceforth, we assume that $g_0,g_1\in \DISJ\cup\NAND\cup\DG$. 

Let us consider $g_0$ first. If $g_0$ belongs to $\DISJ\cup\NAND-\DG$, then Proposition \ref{DISJ-NAND-to-EQ} yields $EQ_{d}\leq_{con}^{+1}\{g_0,u,\Delta_0,\Delta_1\}$ for a certain constraint $u\in\UU\cap\NZ$. 
Since $g_0\leq_{con}^{+0}\{f,\Delta_0\}$, we reach the conclusion   that  $EQ_{d}\leq_{con}^{+1}\{f,u,\Delta_0,\Delta_1\}$ by Lemma \ref{T-transitivity}.  This obviously contradicts our assumption. A similar  contradiction is drawn if we exchange the roles of $g_0$ and $g_1$. Therefore, there is only one remaining case $g_0,g_1\in\DG$ to examine.  
By Lemma \ref{DG-arity-reduction}, since $f\not\in\DG$, we immediately obtain a non-degenerate constraint $g$ of arity $k-1$ such that $g\leq_{con}^{+0}\GG\cup\{f\}$ for a certain finite set $\GG\subseteq\UU\cap\NZ$. If this $g$ is actually in $\DISJ\cup\NAND$, then we conclude, as before, that $EQ_d\leq_{con}^{+1}\GG'\cup\{f\}$ for another finite subset $\GG'$ of $\UU$. Since this is a contradiction, it thus follows that $g\not\in \DISJ\cup\NAND\cup\DG$.  The constraint $g$ certainly satisfies the lemma. 
\end{proof}


In the end, we will prove Proposition \ref{outside-DISJ-NAND} by combining 
Proposition \ref{basis-case-not-DISJ} and Lemma \ref{induction-step-not-DISJ}.

\begin{proofof}{Proposition \ref{outside-DISJ-NAND}}
Let $k\geq2$ and let $f$ be any $k$-ary constraint not in $\DISJ\cup\NAND\cup\DG$. 
Our proof proceeds by induction on the airy $k$ of $f$. 

[Basis Case: $k=2$] For this basis case, 
Proposition  \ref{basis-case-not-DISJ} gives the desired 
conclusion of the proposition.

[Induction Case: $k\geq3$] Our goal is to show that  $EQ_{d}\leq_{con}^{+1}\GG\cup\{f\}$ for a certain set $\GG\subseteq\UU$. 
Toward a contradiction, we assume on the contrary that $EQ_d\not\leq_{con}^{+1}\GG\cup\{f\}$ for any finite subset $\GG$ of 
$\UU$. By Lemma \ref{induction-step-not-DISJ}, there is a constraint $g$ of arity $<k$ for which  
$g\not\in\DISJ\cup\NAND\cup\DG$ and $g\leq_{con}^{+0}\GG'\cup\{f\}$ for a certain finite set $\GG'\subseteq\UU\cap\NZ$. 
We apply the induction hypothesis 
to this  $g$ and then obtain $EQ_{d}\leq_{con}^{+1}\GG''\cup\{g\}$ for another finite set $\GG''\subseteq\UU$. Since $g\leq_{con}^{+0}\GG'\cup\{f\}$, $EQ_{d}\leq_{con}^{+1}\GG'\cup\GG''\cup\{f\}$ follows from Lemma \ref{T-transitivity}.  This is clearly a contradiction; therefore, the proposition holds for $f$.  

Moreover, we obtain the second part of the proposition by appealing to 
Lemma \ref{T-construct-bound}, because the above proof can be efficiently simulated.
\end{proofof}

\section{The Dichotomy Theorem}\label{sec:dichotomy}

Throughout the previous sections, we have already established all necessary foundations for our main theorem---Theorem \ref{main-theorem}---on the approximation complexity of complex-weighted bounded-degree Boolean \#CSPs. Here, we re-state this theorem, which has appeared first in Section \ref{sec:introduction}.

\ms
\n{\bf {\em Theorem \ref{main-theorem}}}\hs{1}{\em (rephrased)\hs{2}
Let $d\geq3$ be any degree bound and let $\FF$ be any set of constraints. If $\FF\subseteq \ED$,  then $\sharpcspstar_{d}(\FF)$ belongs to $\fp_{\complex}$. Otherwise, $\#\mathrm{SAT}_{\complex} \APreduces \sharpcspstar_{d}(\FF)$. 
}
\ms

This theorem is an immediate consequence of our key claim, 
Proposition \ref{equivalence-EQ}, 
which directly bridges between unbounded-degree \#CSPs and 
bounded-degree \#CSPs, when unary constraints are freely available. 
Once the claim is proven, 
the theorem follows from the dichotomy theorem (stated in Section \ref{sec:introduction}) of Yamakami \cite{Yam10}.  Now, we aim at proving  Proposition \ref{equivalence-EQ}.

\ms
\n{\bf {\em Proposition \ref{equivalence-EQ}}}\hs{1}{\em (rephrased)\hs{2}
For any index $d\geq3$ and for any constraint set $\FF$,  $\sharpcspstar(\FF) \APequiv \sharpcspstar_{d}(\FF)$. 
}
\s

\begin{proof}
Let $d$ be any index at least $3$.  Obviously, it holds that $\sharpcspstar_{d}(\FF)\APreduces \sharpcspstar(\FF)$. 
It thus suffices to show the opposite direction 
of this AP-reduction. For convenience, set $\FF'=\FF-\EQ$. 

Let us consider the case where $\FF$ satisfies 
 $\FF\subseteq \ED$.  Lemma \ref{AF-ED-to-FP} directly shows that $\sharpcspstar_{d}(\FF)\in\fp_{\complex}$. Since $\sharpcspstar(\FF)$ 
is also in $\fp_{\complex}$ \cite{Yam10}, $\sharpcspstar(\FF)\APequiv \sharpcspstar_{d}(\FF)$ follows immediately. 
Hereafter, let us assume that  $\FF\nsubseteq\ED$. 
Note that Lemma \ref{F-reduced-to-EQ} helps us AP-reduce  
$\sharpcspstar(\FF)$ to $\sharpcspstar_{2}(\EQ\|\FF')$. 
Now, we want to prove that $\sharpcspstar_{2}(\EQ\|\FF')$ 
is AP-reducible to $\sharpcspstar_{d}(f,\FF')$ for an appropriate constraint $f\in\FF$. This leads us to the conclusion that  
$\sharpcspstar_{2}(\EQ\|\FF')\APreduces \sharpcsp_{d}(\FF)$ since 
$\{f\}\cup\FF'\subseteq \FF$.  

Let us consider the case where either $\FF\subseteq \DISJ \cup \NAND$. 
Since $\FF\nsubseteq \ED$ implies $\FF\nsubseteq \DG$, there exists a constraint $f$ in $\DISJ\cup\NAND-\DG$. The arity of $f$ should be at least $2$ since $f\not\in\DG$. To this $f$,  
we apply Proposition \ref{DISJ-NAND-to-EQ} and then obtain    $\sharpcspstar_{2}(\EQ\|\FF')\APreduces \sharpcsp_{3}(f,\FF')$. 
Since $d\geq3$, we conclude that  $\sharpcspstar_{2}(\EQ\|\FF')\APreduces \sharpcspstar_{d}(f,\FF')$.
The remaining case is that $\FF$ is not included in 
$\DISJ\cup\NAND\cup\DG$. Now, we choose a constraint $f\in\FF$ 
that does not belong to $\cup\DISJ\cup\NAND\cup\DG$. 
Such a constraint can be handled by 
Proposition \ref{outside-DISJ-NAND}. We thus obtain  $\sharpcspstar_{2}(\EQ\|\FF')\APreduces \sharpcspstar_{3}(f,\FF')$, 
which immediately implies  $\sharpcspstar_{2}(\EQ\|\FF)\APreduces \sharpcspstar_{d}(f,\FF')$. This completes the proof. 
\end{proof}

Proposition \ref{equivalence-EQ} is a consequence of the powerful expressiveness of complex-weighted free unary constraints. When free unary constraints are limited to Boolean, Dyer \etalc~\cite{DGJR09} showed a similar proposition only under the assumption that every Boolean constraint in $\FF$ is ``3-simulatable.'' 

Now, Theorem \ref{main-theorem} is immediate from Proposition \ref{equivalence-EQ}.  

\begin{proofof}{Theorem \ref{main-theorem}}
Let $d\geq3$. If  $\FF\subseteq\ED$ holds, 
then $\sharpcspstar_{d}(\FF)$ belongs to 
$\fp_{\complex}$ by Lemma \ref{AF-ED-to-FP}. 
When $\FF\nsubseteq\ED$, as noted in Section \ref{sec:introduction}, it was shown in \cite{Yam10} that $\#\mathrm{SAT}_{\complex} \APreduces \sharpcspstar(\FF)$. 
Since Proposition \ref{equivalence-EQ} establishes the AP-equivalence between $\sharpcspstar(\FF)$ and $\sharpcspstar_{d}(\FF)$, 
we can replace $\sharpcspstar(\FF)$ in the above result by $\sharpcspstar_{d}(\FF)$. This clearly gives the desired consequence of the theorem.
\end{proofof}

Another immediate consequence of Proposition \ref{equivalence-EQ} is 
an AP-equivalence between $\sharpcspstar(\FF)$ and a bipartite Holant problem   $\holant(EQ_3|\FF,\UU)$. 
This immediately follows from the proposition and also a known fact that 
degree-$3$ \#CSPs are essentially identical to bipartite Holant problems whose node labels appearing on the left-hand side of input graphs are always restricted to $EQ_3$. To make this paper self-contained, we will include the detailed proof of the AP-equivalence between $\sharpcspstar(\FF)$ and $\holant(EQ_3|\FF,\UU)$. 

\begin{corollary}\label{AP-reduce-to-EQ_3}
For any set $\FF$ of constraints, it holds that 
$\sharpcspstar(\FF) \APequiv \holant(EQ_3|\FF,\UU)$.
\end{corollary}

\begin{proof}
Let $\FF$ be an arbitrary set of constraints. 
Since $\sharpcspstar_{3}(\FF)$ is shorthand for 
$\sharpcsp_{3}(\FF,\UU)$, by Proposition \ref{equivalence-EQ}, it is enough to prove that $\sharpcsp_{3}(\FF,\UU)$ and 
$\holant(EQ_3|\FF,\UU)$ are AP-equivalent. 
Recall from Section \ref{sec:Holant-problem} that  
$\sharpcsp(\GG)$ always coincides with $\holant(\{EQ_k\}_{k\geq1}|\GG)$ 
for any constraint set $\GG$. In particular, $\sharpcsp_{3}(\FF,\UU)$ coincides with $\holant(EQ_1,EQ_2,EQ_3|\FF,\UU)$. Our goal is therefore set to show that $\holant(EQ_1,EQ_2,EQ_3|\FF,\UU)\APreduces \holant(EQ_3|\FF,\UU)$. 

\sloppy Let us consider any bipartite signature grid $\Omega=(G,\FF'_1|\FF'_2,\pi)$ given as an input instance to $\holant(EQ_1,EQ_2,EQ_3|\FF,\UU)$, where $\FF'_1\subseteq\{EQ_1,EQ_2,EQ_3\}$ and $\FF'_2\subseteq\FF\cup\UU$. Moreover, assume that $G=(V_1|V_2,E)$. 
Now, we  will describe how to replace every node labeled $EQ_1$ with another node whose label is $EQ_3$. For any node $v$ having the label $EQ_{1}$ that appears in $V_1$, let $w$ denote any node, adjacent to $v$, whose label is, say, $g\in\FF'_2$. 
Take any bipartite subgraph $G'=(\{v\}|\{w\},E')$, where $E'$ consists of the edge $(v,w)$ and of all dangling edges obtained from the edges 
linking between the node $g$ and any node other than $v$ in $V_1$. 
We then replace this subgraph $G'$ inside $G$ with the following four-node subgraph $\tilde{G}=(\tilde{V}_1|\tilde{V}_2,\tilde{E})$: $\tilde{V}_1$ is  
composed of a node $v'$ labeled $EQ_3$, 
$\tilde{V}_2$ contains three nodes $w_1,w_2,w_3$, one of which is labeled $g$ and the others are labeled $EQ_1$, and $\tilde{E}$ consists of 
three edges $(v',w_i)$ for all $i\in[3]$ and the original dangling 
edges incident on the node $g$. 
Let $\Omega'$ be the bipartite signature grid obtained from $\Omega$ by 
replacing all nodes labeled $EQ_1$ in $V_1$.  Thus, $\Omega'$ is an input instance to $\holant(EQ_2,EQ_3|EQ_1,\FF,\UU)$, which coincides with $\holant(EQ_2,EQ_3|\FF,\UU)$ because of $EQ_1\in\UU$.  
Note that the aforementioned replacement of two subgraphs does not change 
the value of $\holant_{\Omega}$, and therefore  
we obtain $\holant_{\Omega'} = \holant_{\Omega}$. 

Similarly, we can replace $EQ_2$ by $EQ_3$. When all nodes labeled $EQ_1$ and $EQ_2$ are replaced, we then establish the desired AP-reduction from $\holant(EQ_1,EQ_2,EQ_3|\FF,\UU)$ to $\holant(EQ_3|\FF,\UU)$. 
\end{proof}

\section{Cases of Degree 1 and Degree 2}

When the degree bound $d$ is more than two, our main theorem---Theorem \ref{main-theorem}---has given a complete characterization of the approximation complexity of counting problems $\sharpcspstar_{d}(\FF)$ for any constraint set $\FF$.  This has left a question of what the approximation complexity of $\sharpcspstar_{d}(\FF)$ is, when $d$ is less than three.  
We briefly discuss this issue in this section. 
Let us consider the trivial case of degree one. 

\begin{lemma}
For any constraint set $\FF$, $\sharpcspstar_{1}(\FF)$ is 
in $\fp_{\complex}$.
\end{lemma}

\begin{proof}
Let $\Omega=(G,X|\FF',\pi)$ be any given constraint frame for $\sharpcspstar_{1}(\FF)$. Note that all nodes on the left-hand side of the 
undirected bipartite graph $G$ have degree at most one. 
By this degree requirement, 
no two edges in $G$ are incident on the same node on the left-hand side of $G$. In other words, any two constraints in $\FF'$ share no single  variable.  
This makes $\csp_{\Omega}$ equal to a product of all  values  
$\sum_{\sigma} f(\sigma(x_{i_1}),\sigma'(x_{i_2}),\ldots,\sigma'(x_{i_k}))$ for any constraint  $f\in\FF'$ that takes a variable series $(x_{i_1},x_{i_2},\ldots,x_{i_k})$, where ``sum'' is taken over all variable assignments $\sigma:\{x_{i_1},x_{i_2},\ldots,x_{i_k}\}\rightarrow\{0,1\}$. This value can be easily computed from all constraints in $\FF'$ in polynomial time. Therefore, $\sharpcspstar_{1}(\FF)$ belongs to $\fp_{\complex}$. 
\end{proof}

Next, we consider the case of degree two. Earlier,  
Dyer \etalc~\cite{DGJR09} left this case unanswered for unweighted Boolean \#CSPs. For a complex-weighted case, however, it is possible to obtain 
a precise characterization of $\sharpcspstar_{2}(\FF)$'s using a known transformation between degree-$2$ \#CSPs and Holant problems. 
For completeness, we will formally prove that 
$\sharpcspstar_{2}(\FF)$ is indeed AP-equivalent to $\holant(\FF,\UU)$.  
To simplify the description of Holant problems, similar to the notation 
$\sharpcspstar(\FF)$, we succinctly write $\holantstar(\FF)$ for  $\holant(\FF,\UU)$. 

\begin{proposition}
For any constraint set $\FF$, it holds that $\sharpcspstar_{2}(\FF) \APequiv \holantstar(\FF)$. 
\end{proposition}

\begin{proof}
Firstly, we will claim that $\sharpcspstar_{2}(\FF)$ is AP-equivalent to 
$\holant(EQ_2|\FF,\UU)$. Secondly, we will claim that $\holant(\FF)\APequiv \holant(EQ_2|\FF)$. By replacing $\FF$ by $\FF\cup\UU$, we immediately obtain $\holantstar(\FF)\APequiv \holant(EQ_2|\FF,\UU)$. By combining these two claims, the proposition clearly follows.  

(1) The first claim is proven as follows. In the proof of Corollary \ref{AP-reduce-to-EQ_3}, we have actually proven that $\holant(EQ_1,EQ_2,EQ_3|\FF,\UU) \APequiv \holant(EQ_3|\FF,\UU)$. 
A similar argument shows that $\holant(EQ_1,EQ_2|\FF,\UU)$ and 
$\holant(EQ_2|\FF,\UU)$ are AP-equivalent. Since $\sharpcspstar_{2}(\FF)$ is, as shown in Section \ref{sec:Holant-problem}, essentially the same as $\holant(EQ_1,EQ_2|\FF,\UU)$,  we immediately obtain the desired claim. 

(2) For the second claim, we want to establish 
{\em two} AP-reductions between $\holant(\FF)$ and $\holant(EQ_2|\FF)$.  

(i) In the first step, we prove that $\holant(\FF)$ is AP-reducible to 
$\holant(EQ_2|\FF)$. Let $\Omega=(G,\FF',\pi)$ be any signature grid given as an input instance to $\holant(\FF)$ with $G=(V,E)$. Let us define a new bipartite signature grid $\Omega'=(G',\{EQ_2\}|\FF',\pi')$ as follows. 
For each edge $(v,w)$ incident on both nodes $v$ and $w$ in $G$, we add a new node $u$ labeled $EQ_2$ and replace $(v,w)$ by an edge pair  $\{(u,v),(u,w)\}$. Let $V'_1$ denote the set of all such newly added nodes and let $V'_2$ equal $V$. Let $\pi'$ be obtained from $\pi$ by assigning $EQ_2$ to all the new nodes. A new edge set $E'$ is obtained from $E$ by the above replacement. 
Clearly, $G'=(V'_1|V'_2,E')$ forms an undirected bipartite graph. 
It is not difficult to show that $\holant_{\Omega'} = \holant_{\Omega}$.  Therefore, it holds that $\holant(\FF) \APreduces \holant(EQ_2|\FF)$. 

(ii) In the second step, we will show that $\holant(EQ_2|\FF) \APreduces \holant(\FF)$. 
Fundamentally, we do the opposite of (i), starting from a bipartite signature grid $\Omega'$. More precisely, for any node in $V'_1$, which is labeled $EQ_2$, we delete it and replace each edge pair $\{(u,v),(u,w)\}$ by a new edge $(v,w)$. This defines a new signature grid $\Omega$. Since $\holant_{\Omega} = \holant_{\Omega'}$ holds, we obtain an AP-reduction: 
$\holant(EQ_2|\FF) \APreduces \holant(\FF)$. 
\end{proof}

The computational complexity of exactly solving Holant problems $\holantstar(\FF)$ was completely classified by Cai \etalc~\cite{CLX09@,CLX10} under polynomial-time Turing reductions; on the contrary, it is not known that a similar classification holds 
in the case of approximate counting under AP-reductions.   

\bibliographystyle{alpha}

\begin{thebibliography}{Gur91}
{\small

\bibitem{CL08}
J. Cai and P. Lu. constraint theory in holographic algorithms. 
In {\em Proc. of the 19th International Symposium on Algorithms and Computation} (ISAAC 2008), Lecture Notes in Computer Science, Springer, Vol.5369, 
pp.568--579, 2008.  
\vs{-2}

\bibitem{CL07}
J. Cai and P. Lu. Holographic algorithms: from arts to science. 
{\em J. Comput. Syst. Sci.} 77 (2011) 41--61. 
\vs{-2}

\bibitem{CLX09@}
J. Cai, P. Lu, and M. Xia. Holant problems and counting CSP. In {\em 
Proc. of the 41st Annual ACM Symposium on Theory of Computing} (STOC 2009), pp.715--724, 2009. 
\vs{-2}

\bibitem{CLX10}
J. Cai, P. Lu, and M. Xia. Dichotomy for Holant$^*$ problems of Boolean domain. Preprint, 2010.  
\vs{-2}

\bibitem{CH96}
N. Creignou and M. Hermann. Complexity of generalized satisfiability counting problems. {\em Inform. and Comput.} 125 (1996) 1--12.
\vs{-2} 

\bibitem{CKS01}
N. Creignou, S. Khanna, and M. Sudan. {\em Complexity Classification 
of Boolean Constraint Satisfaction Problems}. SIAM Press, 2001.  
\vs{-2}

\bibitem{DF03}
V. Dalmau and D. K. Ford. Generalized satisfiability with limited concurrences per variable: a study through $\Delta$-matroid parity. 
In {\em Proc. of the 28th International Symposium on 
Mathematical Foundations of Computer Science} (MFCS 2003), 
Lecture Notes in Computer Science, Vol.2747, pp.358--367, 2003.  
\vs{-2}

\bibitem{DFJ02}
M. Dyer, A. Frieze, and M. Jerrum. On counting independent sets in sparse graphs. {\em SIAM J. Comput.} 31 (2002) 1527--1541. 
\vs{-2}

\bibitem{DG00}
M. E. Dyer and C. S. Greenhill. The complexity of counting graph homomorphisms. {\em Random Structures and Algorithms}, 17 (2000) 260--289. Corrigendum appeared in {\em Random Structures and Algorithms} 25 (2004) 346--352. 
\vs{-2}

\bibitem{DGGJ04}
M. Dyer, L. A. Goldberg, C. Greenhill, M. Jerrum. The relative complexity of approximating counting problems. {\em Algorithmica} 38 (2004) 471--500.
\vs{-2}

\bibitem{DGJR09}
M. Dyer, L. A. Goldberg, M. Jalsenius, and D. Richerby. The complexity of approximating bounded-degree Boolean \#CSP. In {\em Proc. of the 27th International Symposium on Theoretical Aspects of Computer Science} (STACS 2010), Leibniz International Proceedings in Informatics, pp.323--334, 2010.  
\vs{-2}

\bibitem{DGJ09}
M. Dyer, L. A. Goldberg, and M. Jerrum. The complexity of 
weighted Boolean \#CSP. {\em SIAM J. Comput.} 38 (2009) 1970--1986.
\vs{-2}

\bibitem{DGJ10}
M. Dyer, L. A. Goldberg, M. Jerrum. An approximation trichotomy for Boolean \#CSP. {\em J. Comput. System Sci.} 76 (2010) 267--277.
\vs{-2}

\bibitem{Fed01}
T. Feder. Fanout limitations on constraint systems. {\em Theor. 
Comput. Sci.} 255 (2001) 281--293. 
\vs{-2}

\bibitem{Ko91}
K. Ko. {\em Complexity Theory of Real Functions.} Birkhauser, 
Cambridge, MA, USA, 1991. 
\vs{-2}

\bibitem{Lad75}
R. E. Ladner. On the structure of polynomial time reducibility. 
{\em J. ACM} 22 (1975)  155--171.  
\vs{-2}

\bibitem{Sch78}
T. J. Schaefer. The complexity of satisfiability problems. In 
{\em Proc.  of the 10th ACM Symposium on Foundations of Computer Science} (FOCS'78), pp.216--226, 1978. 
\vs{-2}

\bibitem{Val79}
L. G. Valiant. The complexity of enumeration and reliability problems. 
{\em SIAM J. Comput.} 8 (1979) 410--421.
\vs{-2}

\bibitem{Val06}
L. G. Valiant. Accidental algorithms. In {\em Proc. of the 47th Annual 
IEEE Symposium on Foundations of Computer Science} (FOCS 2006), pp.509--517, 2006.
\vs{-2}

\bibitem{Val08}
L. G. Valiant. Holographic algorithms. {\em SIAM J. Comput.} 37 (2008) 1565--1594.
\vs{-2}

\bibitem{Yam10}
T. Yamakami. Approximate counting for complex-weighted Boolean constraint satisfaction problems. Available at arXiv:1007.0391. An older version appeared in {\em Proc. of the 8th Workshop on Approximation and Online Algorithms} (WAOA 2010), Lecture Notes in Computer Science, Springer, Vol.6534, pp.261--272, 2011.
\vs{-2}

\bibitem{YY99}
T. Yamakami and A. C. Yao. NQP$_{\complex}$=co-C$_{=}$P. 
{\em Inf. Process. Lett.} 71 (1999) 63--69. 

}
\end{thebibliography}

\end{document}